\documentclass{article}

\usepackage{arxiv}

\usepackage[utf8]{inputenc} 
\usepackage[T1]{fontenc}    
\usepackage{hyperref}       
\usepackage{url}  
\usepackage{float}
\usepackage{amsmath,amsthm}
\usepackage{booktabs}       
\usepackage{amsfonts}       
\usepackage{nicefrac}       
\usepackage{microtype}      
\usepackage{lipsum}
\usepackage{xr}
\usepackage{makecell}
\usepackage{graphicx,amssymb}
\usepackage{xcolor}
\usepackage{authblk}
\graphicspath{ {./images/} }
\usepackage{natbib}

\newtheorem{theorem}{Theorem}[section]
\newtheorem{lemma}[theorem]{Lemma}
\newtheorem{proposition}[theorem]{Proposition}
\newtheorem{corollary}[theorem]{Corollary}
\newtheorem{remark}[theorem]{Remark}

\title{Hormonal Regulation of Breast Cancer Incidence Dynamics: A Mathematical Analysis Explaining the Clemmesen’s Hook}

\author[1,*]{Navid Mohammad Mirzaei}
\author[1,2,*]{Wan Yang}
\affil[1]{Department of Epidemiology, Mailman School of Public Health, Columbia University, New York, New York, USA}
\affil[2]{ Herbert Irving Comprehensive Cancer Center (HICCC), Columbia University Irving Medical Center, New York, New York, USA}

\affil[*]{\textit{Corresponding Authors: nm3519@cumc.columbia.edu (NM); wy2202@cumc.columbia.edu (WY)}}

\allowdisplaybreaks
\begin{document}
\maketitle
\begin{abstract}
Clemmesen’s hook refers to a commonly observed slowdown and rebound in breast cancer incidence around the age at menopause. It suggests a shift in the underlying carcinogenic dynamics, but the mechanistic basis remains poorly understood. Building on our previously developed Extended Multistage Clonal Expansion Tumor (MSCE-T) model, we perform a theoretical analysis to determine the conditions under which Clemmesen’s hook would occur. Our results show that Clemmesen’s hook can be quantitatively explained by time-specific changes in the proliferative and apoptotic balance of early-stage mutated cell populations, corresponding to the decline in progesterone levels and progesterone-driven proliferation due to reduced menstrual cycles preceding menopause, and changing dominant carcinogenic impact from alternative growth pathways post-menopause (e.g., adipose-derived
growth signals). In contrast, variation in last-stage clonal dynamics cannot effectively reproduce the observed non-monotonic incidence pattern. Analytical results further demonstrate that midlife incidence dynamics corresponding to the hook are governed primarily by intrinsic proliferative processes rather than detection effects. Overall, this study provides a mechanistic and mathematical explanation for Clemmesen’s hook and establishes a quantitative framework linking hormonal transitions during menopause to age-specific breast cancer incidence curve.

\end{abstract}

\section{Introduction}\label{sec1}
Breast cancer is the most diagnosed cancer and the main cause of cancer mortality among females worldwide \cite{kim2025global}. The age-specific incidence curves of breast cancer often show a distinctive transition around the menopausal transition (approximately ages 45--55): a reduction and rebound in the slope known as \emph{Clemmesen's hook}, first noticed by the Danish epidemiologist Johannes Clemmesen \cite{clemmesen1965statistical}. This “hook”--visible across cohorts and populations--disrupts the monotonic age patterns and suggests an underlying driver that acts differentially around the age at menopause \citep{anderson2010male,gleason2012breast}. Understanding the origin of this inflection is essential, because the changes in breast cancer risk around menopause in response to the menopausal perturbation (e.g. changing hormonal levels) can help reveal breast cancer development mechanisms for guiding breast cancer prevention and interventions.

Recent epidemiological and biological studies point to ovarian hormones—particularly progesterone—as a causal factor of breast cancer \cite{coelingh2023progesterone,kim2025estrogens,an2022progesterone}. The number of ovulatory menstrual cycles during a person's lifetime, rather than cumulative estrogen exposure alone, has emerged as the dominant determinant of breast cancer risk \cite{coelingh2023progesterone}. For each menstrual cycle, progesterone increases during the luteal-phase, stimulating proliferation of mammary stem and progenitor cells through paracrine RANKL and WNT4 signaling, and in turn increases the probability of replication errors and long-term mutational accumulation \cite{coelingh2023progesterone}. 
Large-scale meta-analyses show that breast-cancer risk increases by 5 \% for each year earlier at menarche and 3 \% for each year later at menopause—consistent with longer exposure to ovulatory cycling and consequently to progesterone surges \cite{collaborative2012menarche}. Conditions that suppress or eliminate these cycles (e.g., lactation, or ovarian suppression) have been shown to be protective against breast cancer development, while exogenous progesterone exposure through progestin-based contraception use or combined estrogen–progestin menopausal therapy elevates breast cancer risk \cite{coelingh2023progesterone,antoine2016menopausal}. Whereas earlier studies suggested a direct carcinogenic role for estrogen \cite{clemons2001estrogen,lupulescu1995clinical}, more recent evidence indicates that unlike progesterone, estrogen alone appears to be permissive rather than causal, increasing the risk for estrogen receptor positive (ER-positive) cancer types through upregulating the progesterone signaling or having milder proliferative effects on progenitors (i.e., unlikely to be causal) \cite{kim2025estrogens,coelingh2023progesterone}. Given these new lines of evidence and the changes in progesterone levels around menopause, we hypothesize that changing progesterone levels is a major driver of the Clemmesen’s hook. That is, the age–incidence curve around menopause (between the ages 45 to 55) is due to the decline of menstruation frequency and progesterone exposure which reduces the proliferation of progenitor cells \cite{collaborative2012menarche,kim2025estrogens,coelingh2023progesterone}.

Mathematical models of carcinogenesis provide an analytic tool to examine cancer incidence trends and the underlying mechanisms. Among these, models based on the Multistage Clonal Expansion (MSCE) framework stand out for their ability to capture how mutation accumulation, clonal growth, and malignant transformation jointly shape age-specific incidence curves. Armitage and Doll laid out the foundation by introducing the theory of multistage carcinogenesis in 1954 \cite{armitage2004age}, and later in 1979 and 1981 Moolgavkar et al. improved it by introducing stage-wise clonal expansion dynamics to the model \cite{moolgavkar1979two,moolgavkar1981mutation}. The MSCE model structure includes initiation, promotion, and malignant conversion, and such models have been shown to replicate various age-dependent cancer incidence curves. However, traditional MSCE models are unable to capture non-monotonic trends and some extensions have to be made depending on the desired functionality such as considering time-dependent parameters to account for evolving risk factors \cite{brouwer2018case,meza2008analysis,mohammad2025modeling} or enhancing the model structure by explicitly incorporating detection-related dynamics \cite{mohammad2025modeling}. Clemmesen’s hook represents one such non-monotonic incidence pattern, warranting some model refinements to accurately reproduce its characteristic slowdown and rebound around ages 45--55 in age-specific breast cancer incidence curve.

To test our hypothesis that reduced progesterone levels shape the incidence pattern (i.e., Clemmesen’s hook) around menopause, we perform a rigorous mathematical analysis of the Multistage Clonal
Expansion Tumor (MSCE-T) model, an extended MSCE model we previously developed to distinguish apparent (detection-related) and true increases in cancer risk \cite{mohammad2025modeling}. We perform a theoretical analysis of the MSCE-T model as well as numerical experiments to determine the conditions under which Clemmesen’s hook would occur.  Our theoretical and numerical analyses consistently demonstrate that the observed age-specific transition (i.e. the hook) can arise from biologically plausible shifts in the effective proliferation rates—consistent with the decline in progesterone exposure and altered hormonal profile during the menopausal transition. We also investigate the effect of detection dynamics on the emergence of Clemmesen's hook. Moreover, we describe the biological pathways embedded in the MSCE-T model that are inherently incapable of generating such transitional behavior, thereby clarifying the model's mechanistic scope in generating the 
hook phenomenon and more broadly deriving the lower bound of mutational steps needed to produce the observed breast cancer age-incidence curve. Together, this study provides an additional layer of quantitative evidence supporting the proposed mechanistic explanation for Clemmesen's hook.

\section{Data and the Model}\label{sec2}
\subsection{Breast Cancer Incidence Data and Clemmesen's hook}

\begin{figure}[H]
    \centering
    \includegraphics[scale=0.5]{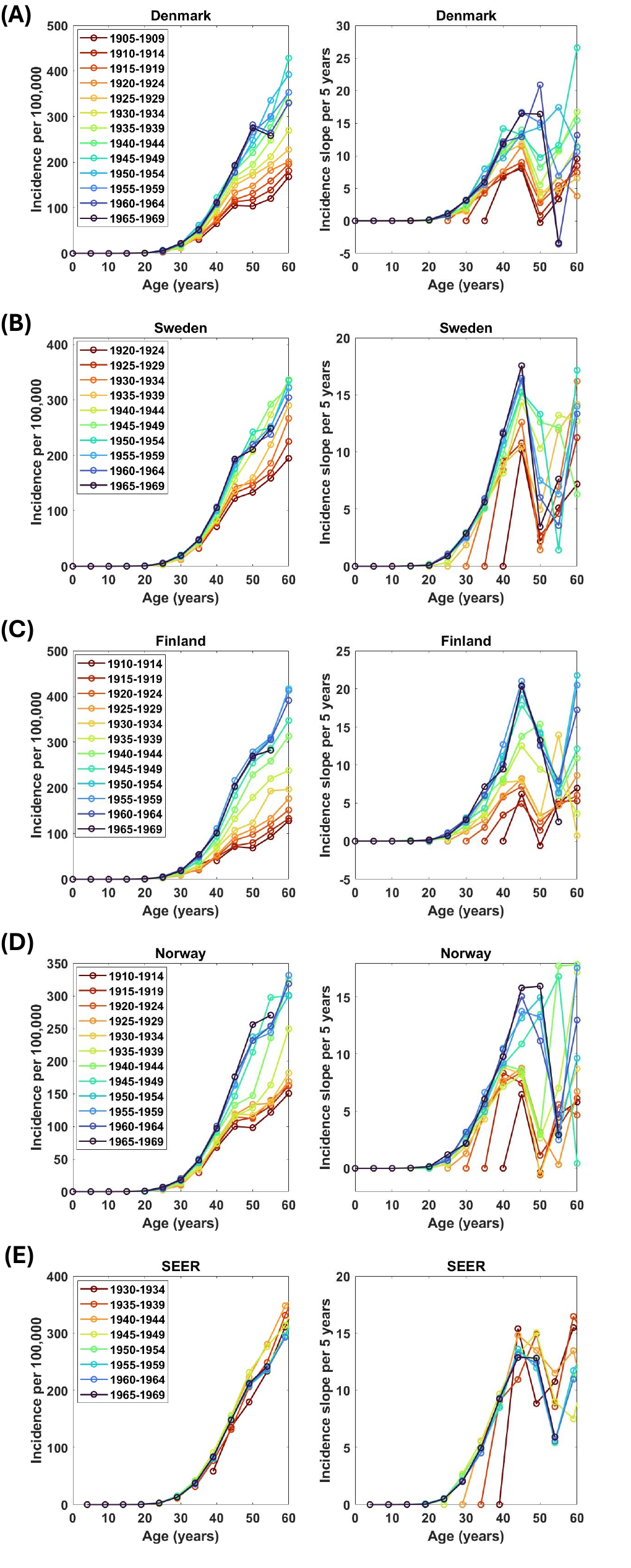}
    \caption{Breast Cancer incidence and incidence slope across different registries. Incidence and incidence slope for (A) Denmark, (B) Sweden, (C) Finland, (D) Norway, and (E) SEER for the United States population.}
    \label{fig1}
\end{figure}

Breast cancer incidence data for Denmark, Sweden, Finland and Norway are sourced from the NORDCAN database \cite{Engholm2010,Laronningen2025}, and data for the U.S. are sourced from the Surveillance, Epidemiology, and End Results (SEER) program \cite{NCI}. 
The NORDCAN database provides the data only in 5-year age groups, while annual data are available from SEER. Using SEER data, we computed the incidence rate by 1-year age interval for each 1-year birth cohort. Note the SEER incidence rates are aggregated to 5-year age interval in Figure \ref{fig1} for clearer visuals, but 1-year-age-specific data are used in our numerical experiments (section 8) when fitting the model. 

Figure \ref{fig1} shows the cancer incidence for the five countries with data, next to the rate of change (i.e. slope). Clemmesen's hook is evident for all five countries and for all birth cohorts. We can see a consistent slowdown and rebound in all of the incidence plots within the 45-55 years window. This is further confirmed by the sudden decrease and increase of the slopes within the same window. 
\raggedbottom

\subsection{The Model}
Considering three rate-limiting driver gene mutations for Breast cancer occurrence \cite{tomasetti2015only,li2018mutation,zhang2005estimating} we extended the classic MSCE model to MSCE-T model by incorporating the detection dynamics \cite{mohammad2025modeling}. A detailed model derivation is provided in our previous work \cite{mohammad2025modeling} and summarized in the Supplemental Materials of this article. The derivation procedure traces the stochastic progression of stem cells as they evolve from normal to malignant states through the sequential accumulation of mutations. Each stage $i$ is modeled as a mutation-birth–death process in which cells may divide (at a rate $\alpha_i > 0$), die (at a rate $\beta_i > 0$), or advance to the next stage via mutation (at a rate $\mu_i > 0$). By incorporating the transition probabilities among these events, the model quantifies the likelihood that a population of malignant cells reaches a detectable size by a given age (i.e. detection dynamics). The model is given by the following ODE system
\begin{align}
\dot x_1 &= \mu_0 N_0\, x_1(x_3-1), \label{eq:x1}\\
\dot x_2 &= -\mu_0 N_0\, x_4, \label{eq:x2}\\
\dot x_3 &= \beta_1-(\alpha_1+\beta_1+\mu_1)x_3+\mu_1 x_3 x_5+\alpha_1 x_3^2, \label{eq:x3}\\
\dot x_4 &= -(\alpha_1+\beta_1+\mu_1)x_4+\mu_1 x_4 x_5+\mu_1 x_3 x_6+2\alpha_1 x_3 x_4, \label{eq:x4}\\
\dot x_5 &= \beta_2-(\alpha_2+\beta_2+\mu_2)x_5+\mu_2 f(t)\,x_5+\alpha_2 x_5^2, \label{eq:x5}\\
\dot x_6 &= -(\alpha_2+\beta_2+\mu_2)x_6+\mu_2 f'(t)\,x_5+\mu_2 f(t)\,x_6+2\alpha_2 x_6 x_5, \label{eq:x6}
\end{align}
with initial condition
\[
\big(x_1(0),x_2(0),x_3(0),x_4(0),x_5(0),x_6(0)\big)=(1,0,1,0,1,0).
\]
and function $f$ which incorporates detection effects
\begin{equation}
f(t)=1-\Big(1-e^{-\alpha_3\,t}\Big)^{M_t-1}. \label{eq:f-def}
\end{equation}
where $\alpha_3 > 0$ is the net proliferation rate of tumor cells and $M_t$ is the number of malignant cells required for detection at age $t$. Function \eqref{eq:f-def} gives the probability of having less than $M_t$ malignant cells at age $t$ assuming a simple birth process. For derivation detail and variable/parameter definitions refer to the supplementary materials.
For convenience set
\begin{equation}\label{eq:a-p-L}
a(t):=1-e^{-\alpha_3\,t}\in(0,1),\quad
p(t):=M_t-1>0,\quad
L(t):=|\ln a(t)|>0.
\end{equation}
The state variables $x_1,x_3$ and $x_5$ are survival probabilities and $h(t):=x_2(t)$ is the hazard of cancer incidence. Also, $x_4 = \dot x_3$ and $x_6=\dot x_5$.



\begin{remark}
    From now on we assume parameters  are piece-wise continuous and bounded functions in time. We are interested in knowing the conditions under which these parameters can produce a midlife transition resembling the Clemmesen's hook for Breast cancer age-specific incidence.
\end{remark}

\begin{remark}
    Since normal mammary cells are in an inactive state until puberty \cite{fu2018foxp1}, the system \eqref{eq:x1}-\eqref{eq:x6} has a zero right hand side (i.e., no carcinogenesis kinetics take place) until the age at menarche ($\sim {13}$ years of age). This does not affect the following proofs, but whenever necessary we will make a note. 
\end{remark}
\section{Positivity, invariance, well-posedness, boundedness}\label{sec:inv}

\begin{lemma}\label{lemma3.1}
Assume $\alpha_i,\beta_i,\mu_i\ge 0$ and $f(t)\in[0,1]$ for all $t\ge 0$. Then the rectangle $[0,1]\times[0,1]$ is forward invariant for the subsystem \eqref{eq:x3} and \eqref{eq:x5}, i.e.\ $x_3(t),x_5(t)\in[0,1]$ for all $t\ge 0$ provided $x_3(0),x_5(0)\in[0,1]$.
\end{lemma}

\begin{proof}
We verify inward-pointing conditions on each edge.
For $x_3=0$, \eqref{eq:x3} gives $\dot x_3=\beta_1\ge 0$; for $x_3=1$,
\[
\dot x_3\big|_{x_3=1}=\beta_1-(\alpha_1+\beta_1+\mu_1)+\mu_1 x_5+\alpha_1
=\mu_1(x_5-1)\le 0 \quad \text{since } x_5\le 1.
\]
Similarly, for $x_5=0$, \eqref{eq:x5} gives $\dot x_5=\beta_2\ge 0$; for $x_5=1$,
\[
\dot x_5\big|_{x_5=1}=\beta_2-(\alpha_2+\beta_2+\mu_2)+\mu_2 f+\alpha_2
=\mu_2(f-1)\le 0 \quad \text{since } f\le 1.
\]
Thus the vector field is inward pointing on all faces of $[0,1]^2$, proving forward invariance.
\end{proof}

\begin{lemma}\label{lem:wp}
Assume $f,f'$ are continuous and bounded on compact time intervals and the parameters are piecewise continuous and bounded. Then $x_3,x_5\in[0,1]$ for all $t$, and $x_4,x_6$ remain bounded on finite time intervals, hence $h'(t)=-\mu_0N_0x_4(t)$ is well-defined and bounded. Moreover, the IVP \eqref{eq:x1}--\eqref{eq:x6} admits a unique solution that is continuous on each subinterval where the parameters are continuous.
\end{lemma}

\begin{proof}
The subsystems for $x_4$ and $x_6$ are linear affine with bounded coefficients on any $[0,T]$:
\[
\dot x_6=q(t)\,x_6(t)+\mu_2 f'(t)\,x_5(t),\qquad
\dot x_4=c(t)\,x_4(t)+\mu_1 x_3(t)\,x_6(t),
\]
with
\begin{equation}\label{eq:cq}
c(t)=-(\alpha_1+\beta_1+\mu_1)+\mu_1x_5(t)+2\alpha_1 x_3(t),\ \
q(t)=-(\alpha_2+\beta_2+\mu_2)+\mu_2 f(t)+2\alpha_2 x_5(t).
\end{equation}
Since $x_3,x_5\in[0,1]$ and $f,f'$ are bounded, $c,q$ are bounded. By integrating factor method,
\[
x_6(t)=\int_0^t \Phi_q(t,s)\,\mu_2 f'(s)x_5(s)\,ds,
\quad \Phi_q(t,s):=\exp\!\Big(\int_s^t q(u)\,du\Big),
\]
hence $|x_6(t)|\le C(T)\int_0^T |f'|$. Likewise,
\[
x_4(t)=\int_0^t \Phi_c(t,s)\,\mu_1 x_3(s)x_6(s)\,ds,
\]
and $|x_4(t)|\le B(T)\int_0^T|x_6|$. Therefore $x_4,x_6$ are bounded on $[0,T]$ for each $T$, precluding blow-up and yielding global well-posedness.

The right-hand side (call it $F({\bf x},t)$) is a polynomial in ${\bf x}=(x_1,\ldots,x_6)$ with time-dependent coefficients that are piece-wise continuous and bounded. Hence, for fixed ${\bf x}$ the mapping $t \mapsto F({\bf x},t)$ is measurable in $t$ and continuous in ${\bf x}$ and for fixed $t$ the mapping $x \mapsto F({\bf x},t)$ is $C^{\infty}$ since it is a polynomial in ${\bf x}$. Also, $x_1, x_3, x_5 \in [0,1]$ and we showed earlier that $x_4$ and $x_6$ are bounded. Given that the parameters and $f$ and $f'$ are also bounded it results in the existence of a uniform bound for $F(\bf x,t)$ for all ${\bf x}$ and $t$. Therefore, the Caratheodory conditions are satisfied for the right-hand side which ensures 
the existence of a solution in a neighborhood of the initial condition. If we show that it is local Lipschitz in ${\bf x}$ then the uniqueness of solution is guaranteed \cite{coddington1956theory}.

Now, let $K$ be a compact subset of $\mathbb{R}^6$ and denote by $D_xF(t,x)$ the Jacobian matrix with entries
$[D_xF(t,x)]_{ij}=\partial F_i/\partial x_j(t,x)$.
By the mean value inequality,
for any $x,y\in K$ we have
\begin{equation}\label{eq:mv}
\|F(t,x)-F(t,y)\|
\le \sup_{\zeta\in[x,y]}\|D_xF(t,\zeta)\|\,\|x-y\|,
\end{equation}

We now bound $\|D_xF(t,\zeta)\|$ uniformly for $\zeta\in K$ and $t\in[0,T]$.
Compute the nonzero partial derivatives component-wise from \eqref{eq:x1}--\eqref{eq:x6}:
\begin{align*}
&\partial_{x_1}F_1=\mu_0N_0(x_3-1), &\partial_{x_3}F_1&=\mu_0N_0x_1;\\
&\partial_{x_4}F_2=-\mu_0N_0;\\
&\partial_{x_3}F_3=-(\alpha_1+\beta_1+\mu_1)+\mu_1x_5+2\alpha_1x_3,
 &\partial_{x_5}F_3&=\mu_1x_3;\\
&\partial_{x_4}F_4=-(\alpha_1+\beta_1+\mu_1)+\mu_1x_5+2\alpha_1x_3,
 &\partial_{x_5}F_4&=\mu_1x_4, \\
&\partial_{x_3}F_4=\mu_1x_6+2\alpha_1x_4,
 &\partial_{x_6}F_4&=\mu_1x_3;\\
&\partial_{x_5}F_5=-(\alpha_2+\beta_2+\mu_2)+\mu_2 f(t)+2\alpha_2x_5;\\
&\partial_{x_6}F_6=-(\alpha_2+\beta_2+\mu_2)+\mu_2 f(t)+2\alpha_2x_5,
 &\partial_{x_5}F_6&=\mu_2 f'(t)+2\alpha_2x_6.
\end{align*}
All other partial derivatives are zero. For $\zeta\in K$ we have $|\zeta_i|\le R$, thus
\begin{align*}
|\partial_{x_1}F_1| &\le |\mu_0N_0|\,(R+1), &
|\partial_{x_3}F_1| &\le |\mu_0N_0|\,R,\\
|\partial_{x_4}F_2| &\le |\mu_0N_0|,\\
|\partial_{x_3}F_3| &\le |\alpha_1|+|\beta_1|+|\mu_1|+|\mu_1|R+2|\alpha_1|R, &
|\partial_{x_5}F_3| &\le |\mu_1|\,R,\\
|\partial_{x_4}F_4| &\le |\alpha_1|+|\beta_1|+|\mu_1|+|\mu_1|R+2|\alpha_1|R, &
|\partial_{x_5}F_4| &\le |\mu_1|\,R,\\
|\partial_{x_3}F_4| &\le |\mu_1|\,R+2|\alpha_1|\,R, &
|\partial_{x_6}F_4| &\le |\mu_1|\,R,\\
|\partial_{x_5}F_5| &\le |\alpha_2|+|\beta_2|+|\mu_2|+|\mu_2|\sup_{t\in[0,T]}|f(t)|+2|\alpha_2|R,\\
|\partial_{x_6}F_6| &\le |\alpha_2|+|\beta_2|+|\mu_2|+|\mu_2|\sup_{t\in[0,T]}|f(t)|+2|\alpha_2|R, &
|\partial_{x_5}F_6| &\le |\mu_2|\,\sup_{t\in[0,T]}|f'(t)|+2|\alpha_2|R.
\end{align*}
Hence there exists a constant $C>0$ (depending only on the parameters) such that
\[
\|D_xF(t,\zeta)\|\;\le\; C\big(1+R+\underbrace{\sup_{t\in[0,T]}|f(t)|}_{M_f}+\underbrace{\sup_{t\in[0,T]}|f'(t)|}_{M_{f'}}\big)
\qquad \text{for all } \zeta\in K,\ t\in[0,T].
\]
Plugging this bound into \eqref{eq:mv} gives
\[
\|F(t,x)-F(t,y)\|
\le C\big(1+R+M_f+M_{f'}\big)\,\|x-y\|
\quad \text{for all } x,y\in K,\ t\in[0,T].
\]
Therefore, setting $\ell_K(t)\equiv C(1+R+M_f+M_{f'})$ yields $\ell_K\in L^\infty([0,T])\subset L^1([0,T])$ and
\[
\|F(t,x)-F(t,y)\|\le \ell_K(t)\,\|x-y\|\quad\text{for\ }t\in[0,T],\ \forall x,y\in K.
\]
Now that the Caratheodory and Lipschitz conditions are satisfied the uniqueness of the solution follows.

\end{proof}
\begin{remark}
    Note that since $\alpha_i, \beta_i$ and $\mu_i$ are piece-wise continuous and bounded, $c(t)$ and $q(t)$ from \eqref{eq:cq} are $L_{loc}^{\infty}$; they may have jump discontinuities but remain bounded.
\end{remark}
\section{Volterra Representation and Exact Form for $f'(t)$}\label{sec:volt}

Solving \eqref{eq:x6} and \eqref{eq:x4} by integrating factors (with $x_4(0)=x_6(0)=0$) gives
\[
x_6(t)=\int_0^t e^{\int_s^t q(u)\,du}\,\mu_2 f'(s)\,x_5(s)\,ds,\qquad
x_4(t)=\int_0^t e^{\int_s^t c(u)\,du}\,\mu_1 x_3(s)\,x_6(s)\,ds.
\]
Plugging $x_6$ into $x_4$ and using Fubini/Tonelli on the triangle $0\le r\le s\le t$ to swap the order of integration yields a Volterra transform of the first kind:
\begin{equation}\label{eq:volterra}
x_4(t)=\int_0^t K(t,r)\,f'(r)\,dr,\quad
K(t,r):=\mu_1\mu_2\,x_5(r)\int_r^t x_3(s)\exp\!\Big(\int_r^s (c+q)\,du\Big)\,ds.
\end{equation}
Differentiating with respect to $t$ gives:
\begin{equation}\label{eq:dK}
\partial_t K(t,r)=\mu_1\mu_2\,x_5(r)\,x_3(t)\exp\!\Big(\int_r^t (c+q)\,du\Big)\ \ge 0.
\end{equation}
Since $h'(t)=-\mu_0N_0x_4(t)$,
\begin{equation}\label{eq:hp-hpp}
h'(t)=A\int_0^t K(t,r) f'(r)\,dr,\qquad
h''(t)=A\int_0^t \partial_t K(t,r)\, f'(r)\,dr,\quad A:=-\mu_0N_0<0.
\end{equation}

Let $a,p$ be as in \eqref{eq:a-p-L}. For $t > 0$ one has $a\in(0,1)$ and $L=|\ln a|=-\ln a>0$.
Differentiating $a=1-e^{-\alpha_3 t}$,
\[
\frac{d}{dt}\big(-\alpha_3 t \big)=-(\alpha_3+t \alpha_3'),\quad
a'(t)=e^{-\alpha_3 t}\,(\alpha_3+ t \alpha_3').
\]
Hence
\begin{equation}\label{eq:a-over-a}
\frac{a'(t)}{a(t)}=\frac{e^{-\alpha_3t}}{1-e^{-\alpha_3t}}\big(\alpha_3+t\alpha_3'\big).
\end{equation}
Using $\frac{d}{dt}\,a^{p}=a^{p}\big(p'\ln a+p\frac{a'}{a}\big)$ and $f=1-a^{p}$,
\begin{equation}\label{eq:fprime}
f'(t)=-\,a(t)^{p(t)}\left[p'(t)\,\ln a(t)+p(t)\,\frac{a'(t)}{a(t)}\right]
=-\,e^{-pL}\left[-p' L + p\,\frac{a'}{a}\right].
\end{equation}

\section{Insensitivity of $h(t)$ to scaling $M_t$}\label{sec:scale}

Fix $c>0$ and a midlife window $[t_a,t_b] \subset [0,T]$ with $T$ being some age well after the menopause window (say $T \ge 60$). For brevity write
\[
p_c(t)=c\,p(t)\quad \text{where} \quad p(t)=M_t-1>0.
\]
and let $f_c,h_c$ be associated with $p_c(t)$. Recall that
\[
a(t)=1-e^{-\alpha_3(t)t},\quad L(t)=|\ln a(t)|>0,
\]
and set
\[
m(t):=\min\{p(t),p_c(t)\}=\min\{1,c\}\,p(t),\qquad p_{\min}:=\inf_{r\in[t_a,t_b]} p(r)>0.
\]
Define the midlife parameters
\[
\varepsilon_1:=\sup_{r\in[t_a,t_b]} L(r),\qquad
\varepsilon_2:=\sup_{r\in[t_a,t_b]}\frac{e^{-\alpha_3(r)r}}{1-e^{-\alpha_3(r)r}}\big(\alpha_3(r)+r|\alpha_3'(r)|\big).
\]
Note that $\varepsilon_1$ and $\varepsilon_2$ are very small for reasonably large $r$ (say $r>20$).

\subsection{Pointwise control of $f'_c-f'$}

\vspace{6pt}
\begin{lemma}\label{lem:pc-refined}
Fix $r\in[t_a,t_b]$. With $R(r):=a'(r)/a(r)$, $L(r)=|\ln a(r)|$ and the definitions above, the following bound holds:
\begin{align}
|f'_c(r)-f'(r)|
&\le \frac{|c-1|}{e}\Bigg[\underbrace{\frac{p(r)}{m(r)}\big(|p'(r)|\,L(r)+p(r)|R(r)|\big)}_{(I)}
\;+\;\underbrace{\frac{|p'(r)|}{m(r)}+\frac{p(r)}{L(r)m(r)}|R(r)|}_{(II)}\Bigg].\label{eq:pc-refined}
\end{align}

\end{lemma}

\begin{proof}
From \eqref{eq:fprime} we have,
\[
f'(r) = -a(r)^{p(r)}\big(-p'(r)L+p(r)R(r)\big),
\quad
f'_c(r) = -a(r)^{p_c(r)}\big(-p_c'(r)L+p_c(r)R(r)\big).
\]
\noindent
So, for the fixed $r$ by adding and subtracting some terms and using the triangle inequality we get
\begin{align}
|f'_c-f'|
&\le |a^{p_c}-a^{p}|\cdot\big(|p'|\,L + p|R|\big)
\;+\; a^{p_c}\big(|p_c'-p'|\,L + |p_c-p|\,|R|\big). \label{eq:pc-start}
\end{align}
\noindent
By the mean value theorem, there exists $\theta$ between $p$ and $p_c$ such that
\[
|a^{p_c}-a^{p}| = a^\theta L\,|p_c-p| \le a^{m}\,L\,|p_c-p|.
\]
By setting $x:=L>0$ and maximizing $x\mapsto x e^{-m x}$ at $x=1/m$ we get
\[
a^{m}L \le \frac{1}{e\,m},
\]
Thus
\[
|a^{p_c}-a^{p}| \le \frac{|p_c-p|}{e\,m} = \frac{|c-1|\,p}{e\,m}.
\]
Insert this into the first term of \eqref{eq:pc-start} to bound it by
\[
\underbrace{\frac{|c-1|}{e}\cdot\frac{p}{m}\big(|p'|\,L + p|R|\big)}_{(I)}.
\]

Now let's bound the second term in \eqref{eq:pc-start}. Note $(p_c'-p')=(c-1)p'$, $(p_c-p)=(c-1)p$ and $a^{p_c}\le 1$. Therefore the second term is bounded by
\[
a^{p_c}\Big(|p_c'-p'|L+|p_c-p||R|\Big) \le |c-1|\big(|p'|\,L + p|R|\big).
\]
Now split this $|c-1|(\cdots)$ to extract explicit $1/m$ factors that we will use to produce the \(1/p_{\min}\) term:
\[
|c-1|\big(|p'|\,L + p|R|\big)
=|c-1|\left[\frac{|p'|}{m} \cdot (mL) \;+\; \frac{p}{L\,m}\cdot (L |R|\,m)\right].
\]
Using the bounds $mL\le 1/(e)$ (shown above) and $L\ge 0$ we get the coarse inequalities (we only need upper bounds)
\[
\frac{|p'|}{m}\cdot (mL) \le \frac{|p'|}{m}\cdot \frac{1}{e},
\qquad
\frac{p}{L\,m}\cdot (L|R|\,m) \le \frac{p}{L\,me}|R| 
\]
Collecting and merging constants, we obtain a bound of the form
\[
\text{(second term)} \le \frac{|c-1|}{e}\Big(\frac{|p'|}{m} + \frac{p}{L\,m}|R|\Big),
\]
which is the term denoted (II) in the statement. Summing the two contributions yields \eqref{eq:pc-refined}:
\[
|f'_c-f'|
\le \frac{|c-1|}{e}\left[\frac{p}{m}\big(|p'|L+p|R|\big) + \frac{|p'|}{m}+\frac{p}{L m}|R|\right].
\]
\end{proof}
\noindent
\begin{remark}
Use $m= \min\{1,c\}\,p$, hence $p/m \le \kappa(c):=\max\{1,1/c\}$ and $1/m= 1/(\min\{1,c\}p)$. Also $L\le \varepsilon_1$, $|R|\le \varepsilon_2$. Define $L_{min} := \inf_{[t_a,t_b]} L(r) $. Replace these into \eqref{eq:pc-refined} to get a more specific bound.
\begin{align}
|f'_c-f'| \le \frac{|c-1|}{e}\left[{\kappa(c)}\big(|p'|\varepsilon_1 + p\varepsilon_2\big) \;+\; \frac{1}{\min\{1,c\}p}\Big(|p'| + \frac{p\varepsilon_2}{L_{min}}\Big)\right]. \label{refined:bnd}
\end{align}
which can be expanded as
\[
|f'_c - f'|
\;\le\;
\frac{|c-1|}{e}\Bigg[
\underbrace{\kappa(c)\big(|p'|\varepsilon_1 + p\,\varepsilon_2\big)}_{\text{(I)}}
\;+\;
\underbrace{\frac{|p'|}{\min\{1,c\}\,p}}_{\text{(II)}}
\;+\;
\underbrace{\frac{\varepsilon_2}{\min\{1,c\}\,L_{\min}}}_{\text{(III)}}
\Bigg].
\]
Turn this into the desired integrand by folding (II)–(III) into a single 
$\dfrac{C_2}{p}$ term and keeping (I) as the 
$\,C_1(|p'|\varepsilon_1 + p\varepsilon_2)\,$ part, With
$C_1 := \frac{\kappa(c)}{e}$ (a constant depending only on \(c\)). For (II), use 
\(|p'(r)| \le \sup_{[t_a,t_b]}|p'| =: \mathcal{M}_{p'}\)
to get
\[
\frac{|c-1|}{e}\cdot\frac{|p'|}{\min\{1,c\}\,p}
\;\le\;
|c-1|\;\frac{1}{p}\,
\underbrace{\Big(\frac{\mathcal{M}_{p'}}{e\,\min\{1,c\}}\Big)}_{=:C_{2a}}.
\]
For (III), use \(p(r)\le \sup_{[t_a,t_b]} p =: \mathcal{M}_p\) so that
\[
\frac{\varepsilon_2}{\min\{1,c\}\,L_{\min}}
\;=\;
\frac{\mathcal{M}_p\,\varepsilon_2}{\min\{1,c\}\,L_{\min}}\cdot\frac{1}{\mathcal{M}_p}
\;\le\;
\frac{\mathcal{M}_p\,\varepsilon_2}{\min\{1,c\}\,L_{\min}}\cdot\frac{1}{p}
\;=\;
\frac{C_{2b}}{p},
\]
with
\[
C_{2b}:=\frac{\mathcal{M}_p\,\varepsilon_2}{e\,\min\{1,c\}\,L_{\min}}. 
\]
\noindent
Combining \(C_{2a}\) and \(C_{2b}\) into \(C_2:=C_{2a}+C_{2b}\) and collecting constants gives
\begin{equation}
|f'_c-f'|
\;\le\;
|c-1|\Big[
C_1\big(|p'|\varepsilon_1 + p\,\varepsilon_2\big)
+
\frac{C_2}{p}
\Big]. \label{eq:refined2}
\end{equation}

\end{remark}
\begin{theorem}\label{thm:scale-sharp}
Under the hypotheses of Lemma~\ref{lem:pc-refined} and knowing that $K(t,r)$ is piece-wise continuous and bounded on $[t_a,t_b]^2$, we have the bound
\begin{equation}\label{eq:final-sharp}
\sup_{t\in[t_a,t_b]}|h_c(t)-h(t)| \ \le\ |c-1|\;C\;\Big(\varepsilon_1+\varepsilon_2+\frac{1}{p_{\min}}\Big),
\end{equation}
where the constant \(C\) depends only on the model constants and the window length:
\[
C:=|A|\,\|K\|_{\infty;[t_a,t_b]^2}\,(T-t_0)\,\widetilde C,
\]
and \(\widetilde C\) depends only on \(c\), \(\sup_{[t_a,t_b]}|p'|\) and \(\sup_{[t_a,t_b]}p\) (but not on \(p_{\min}\)).
Hence
\[
\sup_{t\in[t_a,t_b]}|h_c(t)-h(t)| = O\!\Big(|c-1|\,[\varepsilon_1+\varepsilon_2+1/p_{\min}]\Big).
\]
\end{theorem}

\begin{proof}
Start from the Volterra expression for the derivative (valid because $p_c=p$ on $[0,t_a)$ is assumed, or else we localize to the window as explained earlier):
\[
h_c'(t)-h'(t) = A\int_{t_a}^t K(t,r)\,[f'_c(r)-f'(r)]\,dr,
\]
hence, using $|K(t,r)|\le \|K\|_{\infty;[t_a,t_b]^2}$,
\[
\sup_{t\in[t_a,t_b]}|h_c'(t)-h'(t)| \le |A|\,\|K\|_{\infty;[t_a,t_b]^2}\int_{t_a}^{t_b} |f'_c(r)-f'(r)|\,dr.
\]
Integrate once in time:
\[
|h_c(t)-h(t)| = \Big|\int_{t_a}^t (h_c'(s)-h'(s))\,ds\Big|
\le (t_b-t_a)\sup_{s\in[t_a,t_b]}|h_c'(s)-h'(s)|.
\]
Combining,
\[
\sup_{t\in[t_a,t_b]}|h_c(t)-h(t)|
\le |A|\,\|K\|_{\infty;[t_a,t_b]^2}\,(t_b-t_a)\int_{t_a}^{t_b} |f'_c(r)-f'(r)|\,dr.
\]
Now apply the refined pointwise bound \eqref{eq:refined2} inside the integral:
\[
\int_{t_a}^{t_b} |f'_c-f'|
\le |c-1|\int_{t_a}^{t_b} \Big[ C_1\big(|p'|\varepsilon_1 + p\varepsilon_2\big)+\frac{C_2}{p}\Big]\,dr.
\]
Therefore
\[
\int_{t_a}^{t_b} |f'_c-f'|
\le |c-1|\,(t_b-t_a)\left[C_1\Big(\mathcal{M}_{p'}\varepsilon_1 + \mathcal{M}_p \ \varepsilon_2\Big) + \frac{C_2}{p_{\min}}\right].
\]
Substitute into the previous inequality for $\sup|h_c-h|$:
\[
\sup_{t\in[t_a,t_b]}|h_c(t)-h(t)|
\le |c-1|\,|A|\,\|K\|_{\infty}\,(t_b-t_a)^2\left[C_1\Big(\mathcal{M}_{p'}\varepsilon_1 + \mathcal{M}_p \ \varepsilon_2\Big) + \frac{C_2}{p_{\min}}\right].
\]
Now absorb the 
$(t_b-t_a)$ factors and $\mathcal{M}_p,\;\mathcal{M}_{p'}$ into the constant \( C\) to obtain the stated form \eqref{eq:final-sharp}:
\[
\sup_{t\in[t_a,t_b]}|h_c(t)-h(t)| \le |c-1|\;C\;\Big(\varepsilon_1+\varepsilon_2+\frac{1}{p_{\min}}\Big).
\]

\end{proof}
\noindent
{\bf Caution:} Notice that the term $M_p \ \varepsilon_2$ might be problematic generally (i.e., very large). However, under the circumstances of this study the term is small. Given that we are interested in midlife window of $[45,55]$ relevant to the 
Clemmesen's hook effect, $\varepsilon_2$ is at most $O(e^{-45})$ if not smaller. Even for the supremum of the number of tumor cells at diagnosis within the same window (i.e., $\mathcal{M}_p$) which for breast cancer is at most $O(10^8)$, the product is still very small. 

\subsection{Insensitivity of $h(t)$ to scaling $\alpha_3$}
\begin{theorem}
\label{thm:alpha3-window}
Let $[t_a,t_b]\subset(0,T]$ be a midlife window and let $p(t)=M_t-1>0$ be fixed.
Assume $\alpha_3,\widetilde{\alpha}_3\in C^1([0,T])$ are bounded and coincide
outside the window.
Let $f,h$ and $\widetilde f,\widetilde h$ denote the functions and hazards associated
with the same initial data and model parameters as in \eqref{eq:x1}–\eqref{eq:x6},
but using $\alpha_3$ and $\widetilde{\alpha}_3$ respectively in the definition of
$a(t)$ and $f(t)$. Let $a,L,R$ be defined as in the beginning of the section and similarly $\widetilde a,\widetilde L,\widetilde R$ with $\widetilde{\alpha}_3$. Define $\varepsilon_1, \varepsilon_2$ as before
\[
\varepsilon_1:=\sup_{r\in[t_a,t_b]}L(r),\qquad
\varepsilon_2:=\sup_{r\in[t_a,t_b]}\frac{e^{-\alpha_3(r)r}}{1-e^{-\alpha_3(r)r}}
\bigl(\alpha_3(r)+r|\alpha_3'(r)|\bigr).
\]

\noindent
Then there exists a constant $C>0$, depending only on the model parameters,
$\|p\|_{L^\infty([t_a,t_b])}$, $\|p'\|_{L^\infty([t_a,t_b])}$, the window length
$(t_b-t_a)$ and uniform bounds on $\alpha_3,\alpha_3'$, but \emph{not} on
$\varepsilon_1,\varepsilon_2$, such that
\begin{equation}
\label{eq:alpha3-h-bound}
\sup_{t\in[t_a,t_b]}|\widetilde h(t)-h(t)|
\;\le\;
C\,(\varepsilon_1+\varepsilon_2)\,
\Bigl(
\|\widetilde{\alpha}_3-\alpha_3\|_{L^\infty([t_a,t_b])}
+
\|\widetilde{\alpha}_3'-\alpha_3'\|_{L^\infty([t_a,t_b])}
\Bigr).
\end{equation}
In particular, on a midlife window where $\varepsilon_1,\varepsilon_2\ll1$
(i.e.\ when $t$ is large and $\alpha_3$ varies slowly), even order-one changes in
$\alpha_3$ restricted to $[t_a,t_b]$ induce only negligible changes in the hazard
$h(t)$ on that window.
\end{theorem}

\begin{proof}
We first obtain a Lipschitz bound for $f'(t)$ with respect to $\alpha_3$ and
$\alpha_3'$ on the midlife window, then propagate it to $h$ via the Volterra
representation.
Recall from \eqref{eq:fprime} that
\[
f'(t)
= -a(t)p(t)\bigl[p'(t)L(t)+p(t)R(t)\bigr]
= -e^{-p(t)L(t)}\bigl[-p'(t)L(t)+p(t)R(t)\bigr].
\]
For each fixed $t$, the quantity $f'(t)$ depends on $\alpha_3(t)$ and
$\alpha_3'(t)$ only through the composite map
\[
(\alpha_3(t),\alpha_3'(t))\longmapsto a(t)\longmapsto L(t),R(t)
\longmapsto f'(t),
\]
with $p,p'$ fixed. Since all parameters are bounded and $a(t)\in(0,1)$,
this map is $C^1$ in $(\alpha_3,\alpha_3')$ on a compact set (the range of
$\alpha_3,\alpha_3'$ on $[t_a,t_b]$). Hence there exist partial derivatives
$\partial_{\alpha_3}f'(t)$, $\partial_{\alpha_3'}f'(t)$ that are continuous and
bounded on $[t_a,t_b]$.

Differentiate $f'$ with respect to $\alpha_3$, keeping
$p,p'$ fixed. Writing $E(t):=e^{-pL}$ for simplicity, we get
\[
f'(t)=E(t)\bigl[p'(t)L(t)-p(t)R(t)\bigr],
\]
and thus
\[
\partial_{\alpha_3}f'(t)
= (\partial_{\alpha_3}E)\,\bigl[p'L-pR\bigr]
+ E\Bigl[p'\,\partial_{\alpha_3}L - p\,\partial_{\alpha_3}R\Bigr].
\]
A similar result holds for $\partial_{\alpha_3'}f'(t)$, with
$\partial_{\alpha_3'}L$ and $\partial_{\alpha_3'}R$ in place of
$\partial_{\alpha_3}L$, $\partial_{\alpha_3}R$.
By definition of $E$ and $L$,
\[
\partial_{\alpha_3}E
= -p\,E\,\partial_{\alpha_3}L.
\]
Moreover, from $L=-\ln a$ and $a=1-e^{-\alpha_3 t}$ we have
\[
\partial_{\alpha_3}L(t)
= -\frac{1}{a(t)}\,\partial_{\alpha_3}a(t)
= -\frac{t e^{-\alpha_3(t)t}}{1-e^{-\alpha_3(t)t}},
\]
so $|\partial_{\alpha_3}L(t)|= t\,\frac{e^{-\alpha_3 t}}{1-e^{-\alpha_3 t}}$.
Similarly,
\[
R(t)=\frac{a'(t)}{a(t)}
=\frac{e^{-\alpha_3(t)t}}{1-e^{-\alpha_3(t)t}}
\bigl(\alpha_3(t)+t\alpha_3'(t)\bigr),
\]
one obtains (by chain-rule) the following bounds
\[
|\partial_{\alpha_3}R(t)|+|\partial_{\alpha_3'}R(t)|
\;\le\;C_0\,
\frac{e^{-\alpha_3(t)t}}{1-e^{-\alpha_3(t)t}}
\bigl(\alpha_3(t)+t|\alpha_3'(t)|\bigr),
\]
where $C_0>0$ depends only on uniform bounds for $\alpha_3,\alpha_3'$ and $t$
on $[t_a,t_b]$. By the definition of $\varepsilon_2$, this implies
\[
|\partial_{\alpha_3}R(t)|+|\partial_{\alpha_3'}R(t)|
\;\le\;C_0\,\varepsilon_2
\quad\text{for all }t\in[t_a,t_b].
\]

\noindent
On the other hand, $|L(t)|\le\varepsilon_1$ and
$|R(t)|\le\varepsilon_2$ on $[t_a,t_b]$ by definition of $\varepsilon_1,\varepsilon_2$,
and $0<E(t)\le1$. Combining these bounds in the expression for
$\partial_{\alpha_3}f'(t)$ and the similar expression for
$\partial_{\alpha_3'}f'(t)$, we obtain
\[
|\partial_{\alpha_3}f'(t)| + |\partial_{\alpha_3'}f'(t)|
\;\le\;C_1(\varepsilon_1+\varepsilon_2)
\quad\text{for all }t\in[t_a,t_b],
\]
where $C_1>0$ depends only on $\|p\|_{L^\infty}$, $\|p'\|_{L^\infty}$,
bounds on $\alpha_3,\alpha_3'$ and the window $[t_a,t_b]$, but not on
$\varepsilon_1,\varepsilon_2$.

Let $\delta\alpha_3:=\widetilde{\alpha}_3-\alpha_3$ and
$\delta\alpha_3':=\widetilde{\alpha}_3'-\alpha_3'$, and let
$\delta f'(t):=\widetilde f'(t)-f'(t)$. For each fixed $t\in[t_a,t_b]$, the
mean value theorem in $\mathbb{R}^2$ applied to the $C^1$ map
$(\alpha_3,\alpha_3')\mapsto f'(t)$ yields
\[
|\delta f'(t)|
\;\le\;
\bigl(|\partial_{\alpha_3}f'(t_\theta)|
     +|\partial_{\alpha_3'}f'(t_\theta)|\bigr)
\bigl(|\delta\alpha_3(t)|+|\delta\alpha_3'(t)|\bigr)
\]
for some intermediate values of the arguments (which remain in the same bounded
set). Using the bound we got earlier, we obtain
\[
|\delta f'(t)|
\;\le\;
C_1(\varepsilon_1+\varepsilon_2)
\bigl(|\delta\alpha_3(t)|+|\delta\alpha_3'(t)|\bigr),
\quad t\in[t_a,t_b].
\]
Taking the supremum over $t\in[t_a,t_b]$ yields
\begin{equation}
\label{eq:delta-fprime}
\sup_{t\in[t_a,t_b]}|\widetilde f'(t)-f'(t)|
\;\le\;
C_1(\varepsilon_1+\varepsilon_2)
\Bigl(
\|\widetilde{\alpha}_3-\alpha_3\|_{L^\infty([t_a,t_b])}
+
\|\widetilde{\alpha}_3'-\alpha_3'\|_{L^\infty([t_a,t_b])}
\Bigr).
\end{equation}

\noindent
By the Volterra representation \eqref{eq:hp-hpp} and the assumption
that $\alpha_3=\widetilde{\alpha}_3$ on $[0,t_a)$ (so the kernels $K$ coincide),
we have for $t\in[t_a,t_b]$:
\[
h'(t)-\widetilde h'(t)
= A\int_{t_a}^t K(t,r)\,\bigl(f'(r)-\widetilde f'(r)\bigr)\,dr,
\]
hence
\[
|h'(t)-\widetilde h'(t)|
\;\le\;
|A|\,\|K\|_{\infty;[t_a,t_b]^2}
\int_{t_a}^{t_b}|\widetilde f'(r)-f'(r)|\,dr.
\]
Taking the supremum over $t\in[t_a,t_b]$,
\[
\sup_{t\in[t_a,t_b]}|h'(t)-\widetilde h'(t)|
\;\le\;
|A|\,\|K\|_{\infty;[t_a,t_b]^2}\,(t_b-t_a)\,
\sup_{r\in[t_a,t_b]}|\widetilde f'(r)-f'(r)|.
\]
Integrating once more in time,
\[
|\widetilde h(t)-h(t)|
=
\Bigl|\int_{t_a}^t\bigl(\widetilde h'(s)-h'(s)\bigr)\,ds\Bigr|
\;\le\;
(t_b-t_a)\,\sup_{s\in[t_a,t_b]}|h'(s)-\widetilde h'(s)|,
\]
so that
\[
\sup_{t\in[t_a,t_b]}|\widetilde h(t)-h(t)|
\;\le\;
|A|\,\|K\|_{\infty;[t_a,t_b]^2}\,(t_b-t_a)^2\,
\sup_{r\in[t_a,t_b]}|\widetilde f'(r)-f'(r)|.
\]
Finally, substitute the bound \eqref{eq:delta-fprime} into this inequality and
combine the factors $|A|$, $\|K\|_{\infty;[t_a,t_b]^2}$ and
$(t_b-t_a)^2$ into a single constant $C>0$. This yields
\[
\sup_{t\in[t_a,t_b]}|\widetilde h(t)-h(t)|
\;\le\;
C(\varepsilon_1+\varepsilon_2)\,
\Bigl(
\|\widetilde{\alpha}_3-\alpha_3\|_{L^\infty([t_a,t_b])}
+
\|\widetilde{\alpha}_3'-\alpha_3'\|_{L^\infty([t_a,t_b])}
\Bigr),
\]
which is exactly \eqref{eq:alpha3-h-bound}. The “negligible-effect”
follows by the fact that $\varepsilon_1,\varepsilon_2\ll1$ on the midlife window and noting
that the right-hand side is then small even for order-one perturbations of
$\alpha_3$ and $\alpha_3'$ on the window.
\end{proof}
\section{Dip and rebound of $h'(t)$}\label{sec:dip}

We now give sufficient conditions for a \emph{dip} of $h'(t)$ on a midlife window and a \emph{rebound} later, which amounts to a Clemmesen hook-like behavior in $h(t)$.

\subsection{Sufficient conditions for dip and rebound}

Consider the initial value problem \eqref{eq:x1}-\eqref{eq:x6} along with \eqref{eq:f-def}. In the previous section we showed that changes in the detection dynamics localized to the midlife window is not strong enough to produce a realistic transition phase such as Clemmesen's hook. We now show how to obtain such a hook purely by parameter choices.
Recall
\[
\dot x_4 = c(t)\,x_4(t)+ \mu_1 x_3(t)\,x_6(t), 
\]
where
\[
c(t):=-(\alpha_1+\beta_1+\mu_1)+\mu_1 x_5(t)+2\alpha_1 x_3(t)
=\alpha_1\big(2x_3(t)-1\big)-\beta_1+\mu_1\big(x_5(t)-1\big).
\]
and $h'(t)=A\,x_4(t)$ with $A<0$. 
A hook in $h(t)$ means:
\[
\frac{d}{dt}(-x_4)= -\dot x_4 \ \text{is}
\ \begin{cases}
>0 & \text{(pre-window growth)},\\
<0 & \text{(mid-window decline)},\\
>0 & \text{(post-window growth)}.
\end{cases}
\]
Equivalently, we want
\[
\dot x_4 \ \text{to be}\ 
\begin{cases}
<0 & \text{before $t_a$ },\\
>0 & \text{on the window }[t_a,t_b],\\
<0 & \text{after }t_b.
\end{cases}
\tag{$\star$}
\]
Since, in general $f'(t)\le 0$ (given the improvement in detection over time), the inhomogeneous input in 
\[
\dot x_6 = q(t)x_6(t)+\mu_2 x_5(t) f'(t), \qquad q(t):=-(\alpha_2+\beta_2+\mu_2)+\mu_2 f(t)+2\alpha_2 x_5(t)
\]
is nonpositive. With $x_6(0)=0$, the Volterra representation of $x_6$ in section 4 gives $x_6(t)\le 0$. Given that mutational events are rare the mutation rates $\mu_i$ are typically small. Given that $x_3$ and $x_5$ (survival probabilities) get smaller for later ages we can deduce that $|\mu_1 x_3 x_6|$ is relatively small. Then 
\[
\dot x_4 \ =\ c(t)\,x_4(t) \ +\ \underbrace{\mu_1 x_3(t) x_6(4)}_{\le 0}
\ \le\ c(t)\,x_4(t),
\]
and, when $|\,\mu_1 x_3 x_6|\ll |c(t)x_4|$, the sign of $\dot x_4$ is dictated by $c(t)\,x_4$.

At $t=0$, $x_4(0)=0$ and $\dot x_4(0)=\mu_1 x_3(0) x_6(0)=0$. 
If $f'(0)<0$ (or shortly after), then $x_6<0$ for small $t$, hence $\dot x_4\approx \mu_1 x_3 x_6<0$, so $x_4$ becomes negative and $-x_4>0$ corresponding to the increase in the incidence. In early years, when $x_3, x_5 \approx 1$ for $\alpha_1>\beta_1$ we get $c(t)>0$ which makes $c(t) x_4<0$ and consequently $\dot{x}_4<0$. 
These parameters are biologically relevant and bounded. Here as mutations occur in tumor suppressor genes or oncogenes leading to lowered cell death rates ($\beta_i$) or increased proliferation rates ($\alpha_i$) respectively, it is most likely that the mutated cells grow (i.e., $\alpha_1>\beta_1$) following the initial mutation (at the rate $\mu_0$).

Now, we investigate the most reasonable parameter choices that could create a midlife dip and rebound.

\begin{proposition}\label{prop:mu1-lb}
Let $[t_a,t_b]$ be a midlife window. Fix $t\in[t_a,t_b]$ with $x_5(t)<1$ and $x_3(t)\in(0,1)$. A necessary condition for $c(t)<0$ is
\begin{equation}\label{eq:mu1min}
\mu_1(t)\ \ge\  \frac{\beta_1-\alpha_1\,(2 x_3(t)-1)}{1-x_5(t)}.
\end{equation}

\end{proposition}

\begin{proof}
$c(t)<0$ is equivalent to
\[
\alpha_1(2 x_3-1)-\beta_1+\mu_1(x_5-1)<0
\quad\Longleftrightarrow\quad
\mu_1(1-x_5)>\beta_1-\alpha_1(2 x_3-1).
\]
Since $x_5<1$ we may divide by $1-x_5>0$ to obtain \eqref{eq:mu1min}.
\end{proof}

{\bf Note:} Numerically and without intervention when the model is fit to the incidence data $\alpha_1>\beta_1$ but their values are close (see the Numerical Experiment section). Therefore, there is a time $t>0$ such that $2x_3(t)-1 \lessapprox 1$ and $\beta_1-\alpha_1 (2 x_3-1)>0$. After that the difference gets larger and larger. Given that the denominator is also less than 1 for $t>0$ then $\mu_1$ likely requires an extreme scale-up to satisfy the necessary condition.

\begin{lemma}\label{lem:comp-x3}
Let $\mu_1,\widetilde\mu_1$ be such that
\[
\widetilde\mu_1(t)\ \ge\ \mu_1(t)\quad \text{for }t\in[t_a,t_b],\qquad
\widetilde\mu_1(t)=\mu_1(t)\quad \text{for }t\notin[t_a,t_b].
\]
Let $x_3,\widetilde x_3$ be the corresponding solutions with the same initial state at $t=t_a$. Then
\begin{equation}
\widetilde x_3(t)\ \le\ x_3(t)\qquad \text{for all } t\ge t_a. \label{eq:comp}
\end{equation}
\end{lemma}

\begin{proof}
Write
\[
f_3(x_3;t,\mu_1):=\beta_1-(\alpha_1+\beta_1+\mu_1)\,x_3+\mu_1 x_3 x_5+\alpha_1 x_3^2
= \beta_1-(\alpha_1+\beta_1)x_3-\mu_1 x_3(1-x_5)+\alpha_1 x_3^2,
\]
so $\dot x_3=f_3(x_3;t,\mu_1)$. For any fixed $t$ and $x_3\in(0,1]$ with $x_5(t)\in(0,1]$,
\[
\frac{\partial f_3}{\partial \mu_1}(x_3;t,\mu_1)= -\,x_3(1-x_5(t))\ \le\ 0.
\]
Thus $f_3(\cdot;t,\mu)$ is nonincreasing in $\mu_1$ pointwise. Let $y:=\widetilde x_3-x_3$. Then
\[
\dot y \ =\ f_3(\widetilde x_3;t,\widetilde\mu_1)-f_3(x_3;t,\mu_1)
\ =\ \big[f_3(\widetilde x_3;t,\widetilde\mu_1)-f_3(\widetilde x_3;t,\mu_1)\big]\ +\ \big[f_3(\widetilde x_3;t,\mu_1)-f_3(x_3;t,\mu_1)\big].
\]
The first bracket is $\le 0$ on $[t_a,t_b]$ by $\widetilde\mu_1\ge\mu_1$ and equals $0$ outside. Using the mean value theorem in $x_3$ and the fact that the following is bounded on any compact $x_3$-interval (in particular on $[0,1]$)
\[
\frac{\partial f_3}{\partial x_3}(x_3;t,\mu_1)= -(\alpha_1+\beta_1+\mu_1)+\mu_1 x_5+2\alpha_1 x_3
\]
there exists $L>0$ with
$|f_3(\widetilde x_3;t,\mu_1)-f_3(x_3;t,\mu_1)|\le L\,|y|$.
Therefore,
\[
\dot y(t)\ \le\ L\,y(t),\qquad y(t_a)=0,
\]
and Grönwall’s inequality gives $y(t)\le 0$ for all $t\ge t_a$, i.e. \eqref{eq:comp}.
\end{proof}

\begin{lemma}\label{lem:threshold}
For any $t\ge t_b$ and any $\mu_1^{\mathrm{post}}\ge 0$,
\begin{equation}\label{eq:c-upper}
c(t)\ \le\ \alpha_1\big(2x_3(t)-1\big)-\beta_1.
\end{equation}
Hence a necessary condition for $c(t)>0$ is
\begin{equation}\label{eq:threshold}
x_3(t)\ >\ \tau := \frac12+\frac{\beta_1}{2\alpha_1}.
\end{equation}
\end{lemma}

\begin{proof}
Since $x_5(t)\le 1$, we have $\mu_1^{\mathrm{post}}(x_5(t)-1)\le 0$, which implies \eqref{eq:c-upper}. Rearranging $\alpha_1(2x_3-1)-\beta_1>0$ yields \eqref{eq:threshold}.
\end{proof}

\begin{proposition}\label{prop:no-rebound}
Suppose that on $[t_a,t_b]$ we have $\widetilde\mu_1\ge\mu_1$ (relative to a baseline $\mu_1$) that is used to achieve $c<0$ on $[t_a,t_b]$ and that the corresponding solution satisfies
\[
\widetilde x_3(t_b^+)\ \le\ \tau.
\]
Then for every $\mu_1^{\mathrm{post}}\ge 0$ and every $t\ge t_b$ we have $c(t)\le 0$. In particular, lowering $\mu_1$ back to baseline or even below baseline after $t_b$ cannot produce $c(t)>0$ (no rebound).
\end{proposition}

\begin{proof}
By Lemma \ref{lem:comp-x3}, $\widetilde x_3(t)\le x_3^{\mathrm{base}}(t)$ for $t\ge t_a$. At $t_b^+$, $\widetilde x_3(t_b^+)\le\tau$ by assumption. By continuity of $t\mapsto \widetilde x_3(t)$ there exists $\delta>0$ with $\widetilde x_3(t)\le\tau$ on $[t_b,t_b+\delta]$. For such $t$,
Lemma \ref{lem:threshold} yields
\[
c(t)\ \le\ \alpha_1\big(2\widetilde x_3(t)-1\big)-\beta_1\ \le\ \alpha_1(2\tau-1)-\beta_1\ =\ 0.
\]
Hence $c(t)\le 0$ on $[t_b,t_b+\delta]$. Note that $\widetilde x_3(t)$ is a survival probability so it is nonincreasing in time so $\widetilde x_3(t)\le\tau$ for all $t\ge t_b$, then $c(t)\le 0$ for all $t\ge t_b$, which rules out a rebound.
\end{proof}

\noindent
\textbf{Conclusion}: Based on the propositions and lemmas proved above, to begin a dip (i.e., $c(t)<0$) in the midlife window we likely need a large increase in $\mu_1$, which in turn leads to a smaller $x_3$ value. Now to get back to a $c(t)>0$ we need $x_3\le \tau$ to become $x_3 \ge \tau$ but we cannot. So, a dip and rebound cannot be achieved by changing $\mu_1$.

\vspace{0.2in}

Now consider the following piecewise-constant schedule:
\[
\alpha_1(t)=
\begin{cases}
\alpha_1^{\mathrm{base}}-\Delta_\alpha, & t\in[t_a,t_b],\quad \Delta_\alpha>0,\\[2pt]
\alpha_1^{\mathrm{base}}, & t>t_b,
\end{cases}
\qquad
\beta_1(t)=
\begin{cases}
\beta_1^{\mathrm{base}}, & t\le t_b,\\[2pt]
\beta_1^{\mathrm{base}}-\Delta_\beta, & t>t_b,\quad \Delta_\beta>0,
\end{cases}
\]
with $\mu_1$ fixed at baseline.

\begin{proposition}\label{prop:dip-alpha}
Let $\alpha_1^{\mathrm{new}}=\alpha_1^{\mathrm{base}}-\Delta_\alpha$ with $\Delta_\alpha>0$ on $[t_a,t_b]$.
Then for a fixed $t \in [t_a,t_b]$ the change in $c$ satisfies
\[
c_{\mathrm{new}}(t)-c_{\mathrm{base}}(t)
= -\Delta_\alpha\,(2x_3(t)-1).
\]
In particular, on any subinterval where $x_3(t)>\tfrac12$, decreasing $\alpha_1$ strictly decreases $c(t)$.
\end{proposition}

\begin{proof}
Direct substitution:
$c=\alpha_1(2x_3-1)-\beta_1+\mu_1(x_5-1)$, from which
$\Delta c = (\alpha_1^{\mathrm{base}}-\Delta_\alpha-\alpha_1^{\mathrm{base}})(2x_3-1) = -\Delta_\alpha(2x_3-1)$.
\end{proof}

\begin{remark}\label{lem:x3-alpha}
On $[t_a,t_b]$, $\partial f_3/\partial \alpha_1=x_3(x_3-1)\le 0$. Hence, replacing $\alpha_1^{\mathrm{base}}$ by $\alpha_1^{\mathrm{base}}-\Delta_\alpha$ \emph{increases} the right-hand side of $\dot x_3=f_3$ pointwise, and comparison yields
\[
x_3^{(\alpha_1^{\mathrm{base}}-\Delta_\alpha)}(t)\ \ge\ x_3^{(\alpha_1^{\mathrm{base}})}(t)
\qquad (t\in[t_a,t_b]).
\]
Thus the window intervention (through lowering $\alpha_1$) does not reduce $x_3$ at $t_b^+$ relative to baseline. (Similar reasoning as in Lemma \ref{lem:comp-x3}). This is important because if lowering $\alpha_1$ lowered $x_3$ in addition to $c(t)$ then there will not be any chance to achieve $c(t)>0$ after the window by moderate changes in $\beta_1$.
\end{remark}

\begin{proposition}\label{prop:rebound-beta}
At $t_b^+$, change $\beta_1\to \beta_1-\Delta_\beta$ with $\Delta_\beta>0$. Then there exists $\delta>0$ (depending on local Lipschitz bounds of $f_3$) such that
\[
c_{\mathrm{new}}(t)\ >\ 0\qquad\text{for all }t\in(t_b,t_b+\delta].
\]
\end{proposition}

\begin{proof}
Set $\Delta x_3(t):=x_3^{\mathrm{new}}(t)-x_3^{\mathrm{base}}(t)$ for $t\ge t_b$.
Write
\[
\dot{\Delta x_3}
= \underbrace{\big[f_3(x_3^{\mathrm{new}};t,\beta_1)-f_3(x_3^{\mathrm{base}};t,\beta_1)\big]}_{=:A(t)}
 + \underbrace{\big[f_3(x_3^{\mathrm{new}};t,\beta_1-\Delta_\beta)-f_3(x_3^{\mathrm{new}};t,\beta_1)\big]}_{=:B(t)} .
\]
Since $f_3$ is locally Lipschitz in $x_3$, $|A(t)|\le L\,|\Delta x_3(t)|$ for some $L>0$.
Moreover, 
\[
B(t)= -\Delta_\beta\,(1-x_3^{\mathrm{new}}(t)) \ \le\ 0.
\]
Hence
\[
|\dot{\Delta x_3}(t)|
\le L\,|\Delta x_3(t)| + \Delta_\beta \qquad (t\ge t_b), \qquad |\Delta x_3(t_b)|=0.
\]
using the fact that $\frac{d}{dt}|\Delta x_3(t)| \le |\dot{\Delta x_3}(t)|$, we get:
\[
\frac{d}{dt}|\Delta x_3(t)|
\le L\,|\Delta x_3(t)| + \Delta_\beta 
\]
By Grönwall, for $t$ close to $t_b$ we obtain the quantitative bound
\[
|\Delta x_3(t)| \ \le\ \frac{\Delta_{\beta}}{L}\,\big(e^{L(t-t_b)}-1\big)
\]
\noindent
Now compute the change in $c$:
\[
c_{\mathrm{new}}(t)-c_{\mathrm{base}}(t)
= \Delta_{\beta}
\ +\ 2\alpha_1\,\Delta x_3(t).
\]
Therefore, for $t\in[t_b,t_b+\delta]$,
\[
c_{\mathrm{new}}(t)-c_{\mathrm{base}}(t)
\ \ge\ \Delta_\beta\ -\ 2\alpha_1\,|\Delta x_3(t)|
\ \ge\ \Delta_\beta\ -\ 2\alpha_1\,\frac{\Delta_{\beta}}{L}\,\big(e^{L(t-t_b)}-1\big)
\]
Choose $\delta_1>0$ so small that $\frac{2\alpha_1}{L}(e^{L\delta}-1) \le \tfrac12$. Then
\[
c_{\mathrm{new}}(t)-c_{\mathrm{base}}(t)\ \ge\ \tfrac12\,\Delta_\beta \qquad \text{for } t\in(t_b,t_b+\delta_1].
\]
which means 
\[
c_{\mathrm{new}}(t) \ge \tfrac12\,\Delta_\beta+c_{\mathrm{base}}(t). \tag{$\star$}
\]
Now we know that $c_{base}(t)$ is continuous on the interval, so for every $\epsilon>0$ we can find a $\delta_2>0$ such that
\[
|c_{base}(t)-c_{base}(t_b^+)|<\epsilon, \qquad \text{for all} \ t \in (t_b,t_b+\delta_2] \tag{$\star \star$}
\]
At the end of the dip, $c_{base}(t_b^+)$ is around zero or slightly negative. So, pick $\Delta_\beta$ such that
\[
c_{base}(t_b^+)+\frac{1}{2}\Delta_\beta-\epsilon>0 \tag{$\diamond$}
\]
Taking $\delta = \min\{\delta_1,\delta_2\}$, for all $t \in (t_b,t_b+\delta]$  we have
\begin{align*}
    c_{\mathrm{new}}(t) &\ge \tfrac12\,\Delta_\beta+c_{\mathrm{base}}(t) \qquad \text{from ($\star$)}\\
    &\ge c_{base}(t_b^+) - \epsilon +\tfrac12 \qquad \text{from ($\star \star$)}\\
    &> 0 \qquad \text{from ($\diamond$)}
\end{align*}
\end{proof}

\noindent
\textbf{Conclusion:} Proposition \ref{prop:dip-alpha} gives the dip on the midlife window via a modest decrease of $\alpha_1$ (because $2x_3-1<0$ there), Remark \ref{lem:x3-alpha} guarantees no adverse reduction of $x_3$ at the window boundary, and Proposition \ref{prop:rebound-beta} yields a post-window interval with $c>0$ after decreasing $\beta_1$—hence a rebound.\\

\noindent
\textbf{Note:} Here we saw that the change required for the dip in midlife window is easily achieved by reasonable changes in $\alpha_1$. In contrast, for $\mu_1$ the same scale of effect likely requires an unreasonably large increase. Then we showed that by changing $\beta_1$ we can readily foil the dip effect and acquire a rebound. It is worth noting that any changes in $\alpha_1$ and $\beta_1$ that results in decreased $\alpha_1/\beta_1$ ratio in the midlife window and then an increase in the same ratio after the window can reproduce the same effect. For example, increasing $\beta_1$ in the midlife window and increasing $\alpha_1$ after can also produce a similar dip and rebound effect, but we chose the other regime due to more biological relevance related to hormonal changes during and after menopause (see the discussion). \\

In this section we only focused on the effect of $c(t)$ and its parameters $\alpha_1, \beta_1$ and $\mu_1$ on the midlife events. Next, we show that $q(t)$ and its parameters cannot create as strong of an impact.

\section{Why $c(t)$ dominates $q(t)$ in midlife dynamics}

We compare, on a fixed midlife window $W_+=[t_a,t_b]$, the size of the response of $h'(t)$ to small, compactly supported perturbations of $c$ versus $q$. 
Let $\delta c,\delta q\in L^1(W_+)$ be small and supported in $W_+$. We write $\Xi=c+q$ and, for a fixed $(t,r)$, recall
\[
K(t,r)=\mu_1\mu_2\,x_5(r)\int_r^t x_3(s)\exp\!\Big(\int_r^s \Xi(u)\,du\Big)\,ds,
\qquad
h'(t)=A\int_0^t K(t,r)f'(r)\,dr.
\]
We consider Gateaux variations to first order in the perturbation amplitude $\varepsilon$:
\[
\delta_\varepsilon h'(t)
:= \frac{h'[\Xi+\varepsilon(\delta c+\delta q)](t)-h'[\Xi](t)}{\varepsilon},
\quad
\delta h'(t):=\lim_{\varepsilon\to 0}\delta_\varepsilon h'(t),
\]
and similarly for $K$. All bounds below are uniform for $t\in W_+$, and constants may depend on the bounds from Lemma~\ref{lem:wp} and on $W_+$, but not on the specific choice of $\delta c,\delta q$.

\begin{lemma}\label{lem:gateaux-K-correct}
Fix $T>0$. Assume $x_3,x_5\in L^\infty([0,T])$ with $0<m_3\le x_3(t)\le M_3$ and $0<m_5\le x_5(t)\le M_5$ a.e.\ on $[0,T]$, and let $\Xi\in L^\infty([0,T])$.
For $0\le r\le t\le T$ define
\[
K(t,r):=\mu_1\mu_2\,x_5(r)\,I(t,r),\qquad
I(t,r):=\int_{r}^{t} x_3(s)\exp\!\Big(\int_{r}^{s}\Xi(u)\,du\Big)\,ds.
\]
Let $\psi\in L^\infty([0,T])$ (this will play the role $\psi=\delta c+\delta q$) and, for $\varepsilon\in\mathbb{R}$, set $\Xi_\varepsilon:=\Xi+\varepsilon\psi$ and $K_\varepsilon$ the kernel obtained by replacing $\Xi$ with $\Xi_\varepsilon$ in the definition above.
Then for every $0\le r\le t\le T$ the G\^ateaux derivative
\[
\delta K(t,r)\ :=\ \lim_{\varepsilon\to 0}\frac{K_\varepsilon(t,r)-K(t,r)}{\varepsilon}
\]
exists and is given by the formula
\begin{equation}\label{eq:deltaK-correct}
\delta K(t,r)\;=\;\mu_1\mu_2\,x_5(r)\int_{r}^{t}\psi(u)\,
\exp\!\Big(\int_{r}^{u}\Xi\Big)\,
\Big(\int_{u}^{t} x_3(s)\exp\!\Big(\int_{u}^{s}\Xi\Big)\,ds\Big)\,du.
\end{equation}
Equivalently, in terms of $I(t,\cdot)$,
\begin{equation}\label{eq:deltaK-correct-compact}
\delta K(t,r)\;=\;\mu_1\mu_2\,x_5(r)\int_{r}^{t}\psi(u)\,e^{\int_{r}^{u}\Xi}\,I(t,u)\,du.
\end{equation}
Hence the logarithmic derivative has the representation
\begin{equation}\label{eq:log-deriv-weight}
\frac{\delta K}{K}(t,r)\;=\;\int_{r}^{t} w(t,r;u)\,\psi(u)\,du,
\qquad
w(t,r;u):=\frac{e^{\int_{r}^{u}\Xi}\,I(t,u)}{I(t,r)}\ \ (\ge 0).
\end{equation}
\end{lemma}

\begin{proof}
Write $K(t,r)=\mu_1\mu_2\,x_5(r)\,I(t,r)$ with
\[
I(t,r):=\int_{r}^{t} x_3(s)\exp\!\Big(\int_{r}^{s}\Xi\Big)\,ds.
\]
Since $x_5$ does not depend on $\Xi$, it suffices to compute the G\^ateaux derivative of $I$.
For $\Xi_\varepsilon:=\Xi+\varepsilon\psi$ we have
\[
I_\varepsilon(t,r):=\int_{r}^{t} x_3(s)
\exp\!\Big(\int_{r}^{s}\Xi_\varepsilon\Big)\,ds
=\int_{r}^{t} x_3(s)\,e^{\int_{r}^{s}\Xi}\,
\exp\!\Big(\varepsilon\int_{r}^{s}\psi\Big)\,ds.
\]
Fix $(t,r)$ and consider the difference quotient:
\[
\frac{I_\varepsilon(t,r)-I(t,r)}{\varepsilon}
=\int_{r}^{t} x_3(s)\,e^{\int_{r}^{s}\Xi}\,
\frac{e^{\varepsilon\int_{r}^{s}\psi}-1}{\varepsilon}\,ds.
\]
Define $Z(s):=\int_{r}^{s}\psi(u)\,du$. Since $\psi\in L^\infty$,
$|Z(s)|\le \|\psi\|_\infty\,(s-r)\le \|\psi\|_\infty\,(t-r)$ for $s\in[r,t]$.
For $|\varepsilon|\le 1$ the following bound holds
\[
\left|\frac{e^{\varepsilon z}-1}{\varepsilon}\right|
\le e^{|z|}\,|z|,\qquad z\in\mathbb{R},
\]
and implies
\[
\left|x_3(s)\,e^{\int_{r}^{s}\Xi}\,\frac{e^{\varepsilon Z(s)}-1}{\varepsilon}\right|
\le \|x_3\|_\infty\,e^{\|\Xi\|_\infty (s-r)}\,e^{|Z(s)|}\,|Z(s)|
\le C\,e^{C (s-r)}\,(s-r),
\]
with $C:=\max\{\|\Xi\|_\infty,\|\psi\|_\infty,\|x_3\|_\infty\}$ (depending only on the uniform bounds and on $T$). The right-hand side is integrable on $s\in[r,t]$.
Moreover, pointwise in $s$,
\[
\frac{e^{\varepsilon Z(s)}-1}{\varepsilon}\ \longrightarrow\ Z(s)\qquad (\varepsilon\to 0).
\]
By the dominated convergence theorem,
\[
\lim_{\varepsilon\to 0}\frac{I_\varepsilon(t,r)-I(t,r)}{\varepsilon}
=\int_{r}^{t} x_3(s)\,e^{\int_{r}^{s}\Xi}\,Z(s)\,ds
=\int_{r}^{t} x_3(s)\,e^{\int_{r}^{s}\Xi}\,\Big(\int_{r}^{s}\psi(u)\,du\Big)\,ds.
\]
Apply Fubini/Tonelli on the triangle $\{(u,s):\, r\le u\le s\le t\}$:
\[
\int_{r}^{t} x_3(s)\,e^{\int_{r}^{s}\Xi}\,\Big(\int_{r}^{s}\psi(u)\,du\Big)\,ds
=\int_{r}^{t}\psi(u)\,\Big(\int_{u}^{t} x_3(s)\,e^{\int_{r}^{s}\Xi}\,ds\Big)\,du.
\]
Use additivity of the integral in the exponent:
\[
\int_{r}^{s}\Xi=\int_{r}^{u}\Xi+\int_{u}^{s}\Xi
\quad\Longrightarrow\quad
e^{\int_{r}^{s}\Xi}=e^{\int_{r}^{u}\Xi}\,e^{\int_{u}^{s}\Xi}.
\]
Therefore
\[
\int_{u}^{t} x_3(s)\,e^{\int_{r}^{s}\Xi}\,ds
=e^{\int_{r}^{u}\Xi}\,\int_{u}^{t} x_3(s)\,e^{\int_{u}^{s}\Xi}\,ds
=e^{\int_{r}^{u}\Xi}\,I(t,u).
\]
We conclude that
\[
\delta I(t,r):=\lim_{\varepsilon\to 0}\frac{I_\varepsilon(t,r)-I(t,r)}{\varepsilon}
=\int_{r}^{t}\psi(u)\,e^{\int_{r}^{u}\Xi}\,I(t,u)\,du,
\]
which is \eqref{eq:deltaK-correct-compact} upon multiplying by $\mu_1\mu_2\,x_5(r)$.
Finally, dividing both sides by $K(t,r)=\mu_1\mu_2\,x_5(r)\,I(t,r)$ gives
\[
\frac{\delta K}{K}(t,r)
=\int_{r}^{t}\frac{e^{\int_{r}^{u}\Xi}\,I(t,u)}{I(t,r)}\,\psi(u)\,du
=\int_{r}^{t} w(t,r;u)\,\psi(u)\,du,
\]
establishing \eqref{eq:log-deriv-weight}.
\end{proof}

The next two results quantify the relative sizes of the responses to $\delta q$ and to $\delta c$.

\begin{proposition}\label{prop:q-only-corrected}
Fix $T>0$. Assume $x_3,x_5\in L^\infty([0,T])$ with $0<m_3\le x_3\le M_3$ and
$0<m_5\le x_5\le M_5$, and $\Xi=c+q\in L^\infty([0,T])$.
Let $\delta c\equiv 0$ and $\delta q\in L^\infty([0,T])$ be supported in $W_+=[t_a,t_b]\subset[0,T]$. Then for every $t\in W_+$,
\begin{equation}\label{eq:q-only-bound-corrected}
|\delta h'(t)|
\ \le\ |A|\,\|K\|_{\infty;[0,t]^2}\,
\Big(\sup_{u\in[0,t]}\int_u^t |\delta q(s)|\,ds\Big)\,
\int_0^t |f'(r)|\,dr.
\end{equation}
In particular, if $\int_{W_+}|f'|\ll 1$ (i.e., small variations of $f$ in the midlife regime), then
\[
\sup_{t\in W_+}|\delta h'(t)|\ \le\ C_q\,
\Big(\sup_{u\in W_+}\int_u^{t_b}|\delta q(s)|\,ds\Big)\cdot \underbrace{\int_{W_+}|f'(r)|\,dr}_{\ll 1},
\qquad C_q:=|A|\,\|K\|_{\infty;[0,T]^2}.
\]
\end{proposition}

\begin{proof}
By the G\^ateaux derivative (Lemma~\ref{lem:gateaux-K-correct}),
with $\psi=\delta q$ and for $0\le r\le t$,
\[
\frac{\delta K}{K}(t,r)
=\int_{r}^{t}\frac{e^{\int_{r}^{u}\Xi}\,I(t,u)}{I(t,r)}\,\delta q(u)\,du
=\int_{r}^{t} w(t,r;u)\,\delta q(u)\,du,
\]
We can write
\[
e^{\int_{r}^{u}\Xi}\,I(t,u) = e^{\int_{r}^{u}\Xi} \int_{u}^{t} x_3(s)e^{\int_{u}^{s}\Xi}\,ds=\int_{u}^{t} x_3(s)e^{\int_{r}^{s}\Xi}\,ds
\]
Hence, for $r\le u\le t$, $0\le w(t,r;u) \le 1$. This gives us the following inequality:
\[
\left|\frac{\delta K}{K}(t,r)\right|
=\left|\int_r^t w(t,r;u)\,\delta q(u)\,du\right|
\le \int_r^t |\delta q(u)|\,du,
\]
which gives
\[
|\delta K(t,r)|\ \le\ |K(t,r)|\int_r^t |\delta q(u)|\,du.
\]
Finally, take $\sup_{(t,r)\in[0,t]^2}|K(t,r)|=\|K\|_{\infty;[0,t]^2}$ we can obtain:

\[
|\delta h'(t)|\;=\;|A|\left|\int_0^t \delta K(t,r)\,f'(r)\,dr\right|
\ \le\ |A|\,\|K\|_{\infty;[0,t]^2}\,
\Big(\sup_{u\in[0,t]}\int_u^t |\delta q|\Big)\int_0^t |f'(r)|\,dr,
\]
which is \eqref{eq:q-only-bound-corrected}.
\end{proof}

\begin{lemma}\label{lem:var-x4-corrected}
Under the same boundedness hypotheses on $x_3,x_5,\Xi$ as above, let $\delta q\equiv 0$ and $\delta c\in L^\infty([0,T])$ be supported in $W_+$. 
Then the first variation $\delta x_4$ solves the linear inhomogeneous ODE
\[
\delta \dot x_4(t) = c(t)\,\delta x_4(t) + \delta c(t)\,x_4(t),
\qquad \delta x_4(0)=0,
\]
and admits the explicit representation
\begin{equation}\label{eq:var-const-x4-corrected}
\delta x_4(t)\ =\ \int_0^t \Phi_c(t,s)\,\delta c(s)\,x_4(s)\,ds,
\quad \Phi_c(t,s):=\exp\!\Big(\int_s^t c(u)\,du\Big).
\end{equation}
\end{lemma}

\begin{proof}
We start from the original unperturbed ODE for $x_4$ because this is directly incorporated in the ODE for $h'(t)$:
\[
\dot x_4(t) = c(t)\,x_4(t) + \mu_1 x_3(t)\,x_6(t).
\]
Now perturb $c$ by a small parameter $\varepsilon$:
\[
c_\varepsilon(t) := c(t) + \varepsilon \delta c(t).
\]
Denote the corresponding perturbed solution by $x_{4,\varepsilon}(t)$. It satisfies
\begin{equation}\label{eq:perturbed-x4}
\dot x_{4,\varepsilon}(t) 
= c_\varepsilon(t)\,x_{4,\varepsilon}(t) + \mu_1 x_3(t)\,x_{6}(t).
\end{equation}
Note that only $c$ is perturbed, while $q$ and $f$ are unchanged, so $x_3$ and $x_6$ are the same across $\varepsilon$.  
The initial condition remains $x_{4,\varepsilon}(0)=0$.
The first variation of $x_4$ is
\[
\delta x_4(t) := \left.\frac{d}{d\varepsilon}x_{4,\varepsilon}(t)\right|_{\varepsilon=0}.
\]
\noindent
Differentiate \eqref{eq:perturbed-x4} with respect to $\varepsilon$ at $\varepsilon=0$. We compute term by term:

- For the left-hand side:
\[
\frac{d}{d\varepsilon}\dot x_{4,\varepsilon}(t)\Big|_{\varepsilon=0}
= \frac{d}{dt}\Big(\frac{d}{d\varepsilon}x_{4,\varepsilon}(t)\Big|_{\varepsilon=0}\Big)
= \delta \dot x_4(t).
\]

- For the right-hand side:
\[
\frac{d}{d\varepsilon}\big(c_\varepsilon(t)\,x_{4,\varepsilon}(t)\big)\Big|_{\varepsilon=0}
= \frac{d}{d\varepsilon}\Big((c(t)+\varepsilon \delta c(t))\,x_{4,\varepsilon}(t)\Big)\Big|_{\varepsilon=0}.
\]
Expand:
\[
= \delta c(t)\,x_4(t) + c(t)\,\delta x_4(t).
\]

- The second term on the right hand side would go away.

\noindent
Thus the variation $\delta x_4$ satisfies
\[
\delta \dot x_4(t) = c(t)\,\delta x_4(t) + \delta c(t)\,x_4(t)
\]
with $\delta x_4(0)=0$. This is exactly the stated ODE. By variation of constants, the ODE for $\delta x_4$ has solution

\[
\delta x_4(t)
= \int_0^t \Phi_c(t,s)\,\delta c(s)\,x_4(s)\,ds
\]
which is exactly \eqref{eq:var-const-x4-corrected}.
\end{proof}

\begin{proposition}\label{prop:c-only-corrected}
Assume the hypotheses at the start of Proposition~\ref{prop:q-only-corrected}.
Let $\delta q\equiv 0$, and let $\delta c$ be supported in $W_+$. Then for every $t\in W_+$,
\begin{equation}\label{eq:c-only-lb-corrected}
|\delta h'(t)|
\ \ge\ |A|\int_0^t \Phi_c(t,s)\,|\delta c(s)|\,|x_4(s)|\,ds
\ -\ |A|\,\|K\|_{\infty;[0,t]^2}\,
\Big(\sup_{u\in[0,t]}\int_u^t |\delta c(s)|\,ds\Big)\,\int_0^t |f'(r)|\,dr.
\end{equation}
\end{proposition}

\begin{proof}
There are two independent first-order paths:

\emph{(1) Homogeneous $x_4$ path.} By Lemma~\ref{lem:var-x4-corrected},
\[
\delta x_4(t)=\int_0^t \Phi_c(t,s)\,\delta c(s)\,x_4(s)\,ds.
\]
Hence the contribution to $h'$ is
\[
\delta h'_{\rm hom}(t)=-\mu_0N_0\,\delta x_4(t)
=-\mu_0N_0\int_0^t \Phi_c(t,s)\,\delta c(s)\,x_4(s)\,ds.
\]
Taking absolute values gives
\[
|\delta h'_{\rm hom}(t)|\ = |A|\int_0^t \Phi_c(t,s)\,|\delta c(s)|\,|x_4(s)|\,ds.
\]

\emph{(2) Kernel reweighting path.} Independently, the kernel $K$ depends on $\Xi=c+q$, so by Lemma~\ref{lem:gateaux-K-correct} with $\psi=\delta c$ and $0\le w\le 1$,
\[
|h'_{\rm re}(t)| = |A|\left|\int_0^t \delta K(t,r)\,f'(r)\,dr\right|
\le |A| \|K\|_{\infty;[0,t]^2}\,
\Big(\sup_{u\in[0,t]}\int_u^t |\delta c|\Big)\,\int_0^t |f'(r)|\,dr.
\]
so
\begin{align*}
|\delta h'(t) |= |\delta h'_{\rm hom}(t)+h'_{\rm re}(t)| \ge |A|\int_0^t &\Phi_c(t,s)\,|\delta c(s)|\,|x_4(s)|\,ds\\
&- |A| \|K\|_{\infty;[0,t]^2}\,
\Big(\sup_{u\in[0,t]}\int_u^t |\delta c|\Big)\,\int_0^t |f'(r)|\,dr.
\end{align*}

\end{proof}

\begin{theorem}\label{thm:cvq-dominate-corrected}
Assume the boundedness hypotheses above and suppose the midlife variations in $f$ is small:
\[
\int_{W_+}|f'(r)|\,dr\ \le\ \varepsilon_\star\qquad\text{with }\ \varepsilon_\star\ll 1.
\]
Assume further that $x_4$ is bounded away from $0$ on a subset $J\subset W_+$ of positive measure:
\[
\exists\,m_4>0,\ |J|>0\ \text{ such that }\ |x_4(s)|\ge m_4\ \text{ for a.e.\ }s\in J.
\]
Let $\delta q,\delta c$ be supported in $W_+$ with comparable $L^1$ sizes:
$\int_{W_+}|\delta q|\approx\int_{W_+}|\delta c|$.
Then, uniformly for $t\in W_+$,
\[
|\delta h'_q(t)|\ \le\ C_q\left(\sup_{u\in W_+}\int_u^{t_b}|\delta q|\right)\varepsilon_\star,
\]
while
\[
\sup_{t\in W_+}|\delta h'_c(t)|
\ \ge\ C_c\int_{J} |\delta c(s)|\,ds\ -\ C'_c\left(\sup_{u\in W_+}\int_u^{t_b}|\delta c|\right)\varepsilon_\star,
\]
where $C_q=|A|\,\|K\|_{\infty;[0,T]^2}$ and
\[
C_c\ :=\ |A|\,\Big(\inf_{t\in W_+,\,s\in J}\Phi_c(t,s)\Big)\,m_4\ >0,
\]
with $\Phi_c$ bounded below on the compact set $W_+\times J$ by boundedness and local integrability of of $c$. Consequently,
\[
\sup_{t\in W_+}\frac{|\delta h'_q(t)|}{|\delta h'_c(t)|}\ =\ o(1)\qquad \text{as}\quad \varepsilon_\star=\int_{W_+}|f'|\ \to\ 0.
\]
\end{theorem}

\begin{proof}
The $q$-bound is Proposition~\ref{prop:q-only-corrected}. For the $c$-bound, apply
\eqref{eq:c-only-lb-corrected} and restrict the first integral to $J$:
\[
\sup_{t\in W_+}|\delta h'_c(t)|
\ \ge\ |A|\Big(\inf_{t\in W_+,\,s\in J}\Phi_c(t,s)\Big)\,m_4\int_{J} |\delta c(s)|\,ds
\ -\ C'_c\Big(\sup_{u\in W_+}\int_u^{t_b}|\delta c|\Big)\int_{W_+}|f'|.
\]
All constants are finite and positive under the stated hypotheses; the ratio statement follows by dividing the two bounds and letting $\varepsilon_\star\to 0$.
\end{proof}
\begin{corollary}
    In addition to the comparison of the effects between $c$ and $q$ perturbations, these results assert that a less than three-stage carcinogenesis model of breast cancer with the inclusion of detection dynamics is not capable of fully capturing the incidence patterns through reasonably varying parameters.  
\end{corollary}
\begin{remark}
    If a biological parameter $p$ (e.g.\ $\alpha_1$) is perturbed, then $c$ changes and \emph{so do} $x_3,x_5$ (hence $x_6$). Writing the linearized system for $(\delta x_3,\delta x_5,\delta x_6)$ and solving by variation of constants yields $\|\delta x_j\|_{L^\infty(0,T)}\le C_T|\delta p|$ for $j=3,5,6$ with constants depending only on bounded Jacobians. Decomposing
\[
\delta h'=\delta h'_{\rm hom}+\delta h'_{\rm re}+\delta h'_{\rm state}
\]
one obtains
\begin{align*}
|\delta h'_{\rm hom}(t)|
&\ \ge\ |A|\int_0^t \Phi_c(t,s)\,|\partial_p c(s)|\,|x_4(s)|\,ds\cdot |\delta p|,\\
|\delta h'_{\rm re}(t)|
&\ \le\ |A|\|K\|_{\infty;[0,t]^2}\left(\sup_{u\in[0,t]}\int_u^t |\partial_p\Xi(s)|\,ds\right)\left(\int_0^t |f'(r)|\,dr\right)|\delta p|,\\
|\delta h'_{\rm state}(t)|
&\ \le\ C\,|\delta p|\,\Big(1+\int_0^t |f'(r)|\,dr\Big),
\end{align*}
which shows: (i) the homogeneous $x_4$ path retains the same form as before and does \emph{not} carry the small midlife factor; (ii) the kernel reweighting and state paths are at worst proportional to $\int_{W_+}|f'|$ or $O(|\delta p|)$ with bounded constants. Therefore, in the midlife regime, the dominance conclusion of Theorem~\ref{thm:cvq-dominate-corrected} persists qualitatively for parameter-level perturbations when the homogeneous path is present.
\end{remark}

\section{Numerical Experiment}

We conducted a numerical experiment. We fit the extended MSCE model \eqref{eq:x1}-\eqref{eq:x6} to the SEER incidence data for three cohorts: i.e., those born during 1935-1939, 1940-1944 and 1945-1949. We chose these three cohorts to minimize the effect of breast cancer screening on the observed incidence. For reference, the Centers for Disease Control and Prevention's National Center for Health Statistics reports less than 29\% of the population aged 40 and over were screened for these cohorts \cite{NCHS2021}. We use a Hybrid Genetic Algorithm which is a global minimization toolbox in Matlab to minimize the distance of the model hazard (i.e., $x_2=h(t)$) from the data. The fitting process has three steps: (i) we first estimate all parameters by fitting the model to the data up to the onset of the hook at $t_a$;
(ii) using the model output at $t_a$ as the initial condition, we fix all parameters except one ($\theta_i \in \{\mu_0,\mu_1,\mu_2,\alpha_1,\alpha_2,\beta_1,\beta_2\}$) and perform the fitting within the dip transition window $[t_a, t_b]$;
(iii) finally, we take the model output at $t_b$ as the new initial condition, fix all parameters except $\theta_j\in \{\mu_0,\mu_1,\mu_2,\alpha_1,\alpha_2,\beta_1,\beta_2\}$, and complete the fitting for the rebound phase over the interval $[t_b, 60]$. We let the search bounds for parameter estimation in steps ii) and iii) be $[0,2\theta_i]$ and $[0,2\theta_j]$ (i.e., widening the lower bound to 0 and upper bound to twice of the value estimated for the preceding phase). For each cohort, the choices of $t_a$ and $t_b$ within the interval $[45,55]$ are determined by identifying 
(i) the age at which the derivative of the smoothed incidence curve first starts a persistent downward shift, and 
(ii) the age at which the derivative begins to recover upward, respectively.

\begin{figure}[H]
    \centering
    \includegraphics[width=\linewidth]{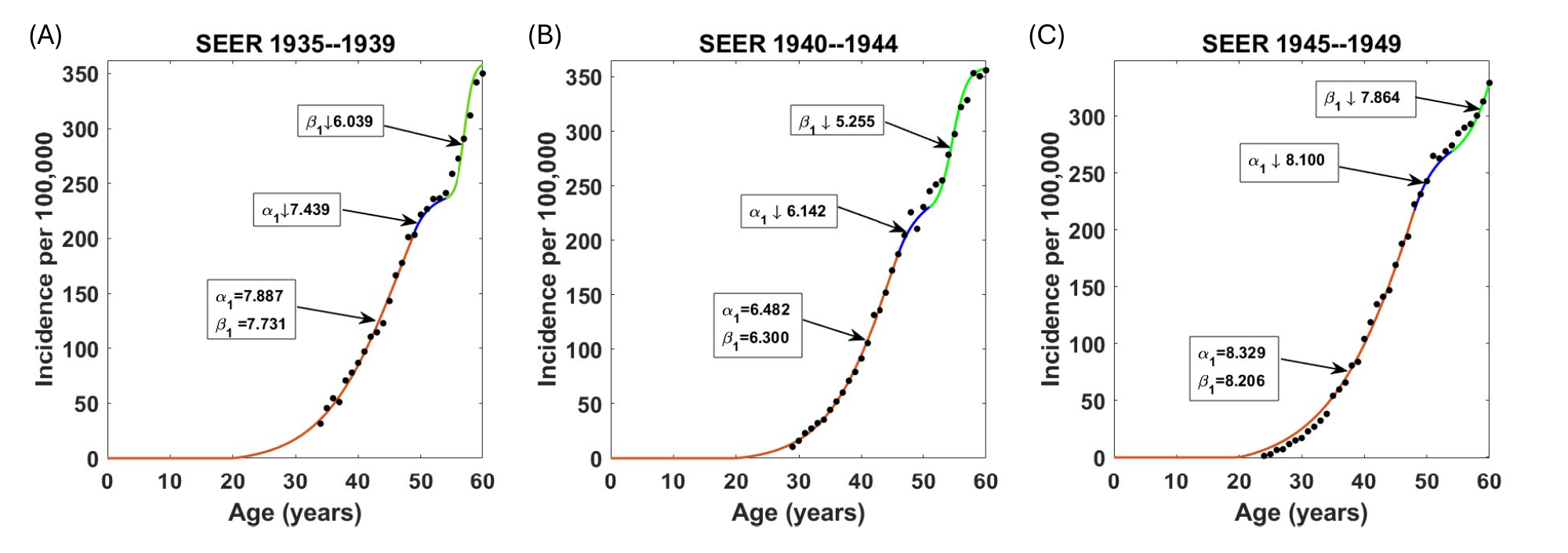}
    \caption{Model fit for SEER incidence data for three cohorts (A) 1935-1939, (B) 1940-1944 and (C) 1945-1949. The orange part of the curves corresponds to the increasing incidence before the occurrence of the Clemmesen's hook. The blue and green parts show the dip and rebound phases, respectively. The midlife window $[t_a,t_b]$ is $[49,54]$ for (A), $[46,51]$ for (B), and $[48,54]$ for (C). The parameters have the unit $\frac{1}{\rm year}$.}
    \label{fig:clemmesen}
\end{figure}

With this, results of our numerical experiments were consistent with the mathematical analysis in showing that generating a hook through variations in $\mu_1$ or in the parameters involved in $q(t)$—namely, $\alpha_2$, $\beta_2$, and $\mu_2$—cannot produce the dip-and-rebound behavior within a reasonable scaling range (i.e., the parameters would need to vary far beyond realistic biological limits to achieve that effect). However, the effect was easily attainable through moderate changes in $\alpha_1$ and $\beta_1$ as shown in Figure \ref{fig:clemmesen}. Equivalently, this refers to changes in the proliferative ratio of first-stage mutated cells ($\frac{\alpha_1}{\beta_1}$).  

As for $\mu_0$, the slowdown and rebound effects were linear, as expected from its direct linear contribution to $x_2$ in \eqref{eq:x2}. Moreover, since $\mu_0$ represents the initiation rate of carcinogenesis—the first mutation event—it governs kinetics that occur early in life, very close to the age at menarche \cite{mirzaei2025estimating,gerstung2020evolutionary}. This suggests that $\mu_0$ is less likely to be influenced by hormonal exposures compared to the other parameters, given the limited number of menstrual cycles before the first mutational event. For a comprehensive numerical comparison of parameter effects and their ability 
to reproduce the hook, please refer to the supplementary materials.

\section{Discussion}

The age-specific incidence of breast cancer exhibits a unique feature known as  Clemmesen’s hook, in which an initial rise in incidence is followed by a temporary slowdown between ages 45 and 55 and then another increase afterwards \cite{clemmesen1965statistical}.The biological mechanisms underlying this pattern remain inconclusive, even though multiple hypotheses have been proposed in several previous studies\cite{abubakar2023host,gleason2012breast,anderson2014many}. In this study we carried out a rigorous mathematical analysis on
an extended multistage clonal expansion model (MSCE-T) representing breast cancer tumor kinetics to investigate the reasons behind this phenomenon.

Our analysis suggests that Clemmesen’s hook likely arises from time-specific changes in the proliferation rate of early mutated clones within a three-stage carcinogenesis framework, occurring midlife around menopause. Specifically, modest reductions in the division rate ($\alpha_1$) or increases in the death rate ($\beta_1$) of first-stage mutated cells are sufficient to reproduce the dip and rebound observed around the age of menopause (i.e., between 45-55). In contrast, changes in other parameters fail to generate a similar pattern unless they are unrealistically large. These findings suggest that the menopausal transition impacts breast cancer development primarily through modulation of clonal expansion rather than imposing mutagenic effects.

The model’s predictions align with biological evidence linking menopause to reduced progesterone-driven proliferation and subsequent activation of growth through other signaling pathways \cite{coelingh2023progesterone,kim2025estrogens,zhao2020cancer}. The transient slowdown in incidence reflects diminished promotion of first-stage mutated cells caused by diminishing exposure to progesterone post-menopause when the ovaries stop producing this hormone (represented by lowered $\frac{\alpha_1}{\beta_1}$ in our model), while the rebound corresponds to continued expansion of existing clones under postmenopausal stimuli, such as adipose-derived growth signals \cite{kim2025estrogens,zhao2020cancer} (corresponding to the recovered $\frac{\alpha_1}{\beta_1}$ in our model).  Our model results support this emerging evidence on the role of progesterone (vs. estrogen) in driving cancer cell proliferation, by specifically connecting the transitions round menopause (Clemmesen’s hook) to model parameters representing the proliferation of early-stage mutated cells around midlife. In addition, our rigorous mathematical analysis helps to rule out other competing pathways (i.e., it is likely due to hormonally regulated changes in clonal expansion kinetics rather than in mutation rate). Further, our numerical analysis fitting the model to the data helps to quantify this impact. 

Mathematically, the Volterra formulation of the MSCE-T model (Section \ref{sec:volt}) separates early-stage carcinogenic dynamics, governed by $c(t)$ in \eqref{eq:cq}, from late-stage and detection-related dynamics, encoded in $q(t)$ and $f(t)$ in \eqref{eq:cq} and \eqref{eq:f-def} respectively. Analytical results show that, within the midlife window, perturbations in $c(t)$ dominate those in $q(t)$, confirming that late-stage dynamics cannot reproduce the hook pattern. Theorems \ref{thm:scale-sharp} and \ref{thm:alpha3-window} further show that the sensitivity of the model hazard $h(t)$ to the tumor detection threshold ($M_t$) and malignant cells proliferation rate ($\alpha_3$) are negligible during the menopausal transition. The analysis also demonstrates that at least three mutational stages are required to reproduce the observed age-incidence trends, consistent with Tomasetti et al. \cite{tomasetti2015only}.

Despite its analytical insight, the model incorporates simplifying assumptions. Parameters $\alpha_i$, $\beta_i$, and $\mu_i$ are treated as deterministic and piecewise-continuous, ignoring stochastic or  variability in hormonal or genetic factors. The model also neglects microenvironmental influences such as oxidative stress, immune activity, and stromal stiffness, which may further modulate proliferation and death rates \cite{hecht2016role,amens2021immune,deng2022biological}. Future extensions that incorporate tissue-level heterogeneity and hormonal trajectories could enable richer dynamics and improve predictions of age-specific risk.

In summary, the extended MSCE-T model provides a coherent mechanistic and mathematical explanation for Clemmesen’s hook by disentangling detection-related effects from intrinsic biological changes. It demonstrates that biologically plausible, time-specific shifts in the proliferation rate of early mutated clones, consistent with menopausal hormonal alterations, are sufficient to explain the midlife dip and rebound in breast cancer incidence. These results establish a quantitative framework for linking hormonal physiology, multi-stage carcinogenesis, and population-level cancer incidence data.

{\bf Supplementary information}:

Model derivation and supplementary numerical experiments are shared as supplementary materials.

\section*{Declarations}

\begin{itemize}
\item {\bf Funding}: This work was supported by the National Institutes of Health (R01CA257971).
\item {\bf Conflict of interest}: The authors declare no conflict of interest.
\item {\bf Data availability}: The data that support the findings of this study are publicly available from the SEER Program at https://seer.cancer.gov/data-software/ and NORDCAN program at https://nordcan.iarc.fr. 
\item {\bf Code availability}: Codes for parameter estimation are available from https://codeocean.com/capsule/6223220/tree/v1 related to our earlier study \cite{mohammad2025modeling}. Further information
is available from the corresponding authors upon request.
\item {\bf Author contribution}: \\
N.M.: Conceptualization, Methodology, Formal analysis, Software, Visualization, Writing – original draft, review \& editing.\\
W.Y.: Supervision, Data curation, Funding acquisition, review \& editing. \\
All authors read and approved the final manuscript.
\end{itemize}








\bibliographystyle{unsrt}
\bibliography{refs.bib}

@article{kim2025global,
  title={Global patterns and trends in breast cancer incidence and mortality across 185 countries},
  author={Kim, Joanne and Harper, Andrew and McCormack, Valerie and Sung, Hyuna and Houssami, Nehmat and Morgan, Eileen and Mutebi, Miriam and Garvey, Gail and Soerjomataram, Isabelle and Fidler-Benaoudia, Miranda M},
  journal={Nature Medicine},
  pages={1--9},
  year={2025},
  publisher={Nature Publishing Group US New York}
}

@article{clemmesen1965statistical,
  title={Statistical Studies in the Aetiology of Malignant Neoplasms. I. Review and Results.},
  author={Clemmesen, Johannes},
  year={1965}
}

@article{coelingh2023progesterone,
  title={Progesterone from ovulatory menstrual cycles is an important cause of breast cancer},
  author={Coelingh Bennink, Herjan JT and Schultz, Iman J and Schmidt, Marcus and Jordan, V Craig and Briggs, Paula and Egberts, Jan FM and Gemzell-Danielsson, Kristina and Kiesel, Ludwig and Kluivers, Kirsten and Krijgh, Jan and others},
  journal={Breast cancer research},
  volume={25},
  number={1},
  pages={60},
  year={2023},
  publisher={Springer}
}

@article{kim2025estrogens,
  title={Estrogens and breast cancer},
  author={Kim, J and Munster, PN},
  journal={Annals of Oncology},
  volume={36},
  number={2},
  pages={134--148},
  year={2025},
  publisher={Elsevier}
}

@article{an2022progesterone,
  title={Progesterone activates GPR126 to promote breast cancer development via the Gi pathway},
  author={An, Wentao and Lin, Hui and Ma, Lijuan and Zhang, Chao and Zheng, Yuan and Cheng, Qiuxia and Ma, Chuanshun and Wu, Xiang and Zhang, Zihao and Zhong, Yani and others},
  journal={Proceedings of the National Academy of Sciences},
  volume={119},
  number={15},
  pages={e2117004119},
  year={2022},
  publisher={National Academy of Sciences}
}

@article{collaborative2012menarche,
  title={Menarche, menopause, and breast cancer risk: individual participant meta-analysis, including 118 964 women with breast cancer from 117 epidemiological studies},
  author={Collaborative Group on Hormonal Factors in Breast Cancer and others},
  journal={The lancet oncology},
  volume={13},
  number={11},
  pages={1141--1151},
  year={2012},
  publisher={Elsevier}
}

@article{antoine2016menopausal,
  title={Menopausal hormone therapy use in relation to breast cancer incidence in 11 European countries},
  author={Antoine, Caroline and Ameye, Lieveke and Paesmans, Marianne and de Azambuja, Evandro and Rozenberg, Serge},
  journal={Maturitas},
  volume={84},
  pages={81--88},
  year={2016},
  publisher={Elsevier}
}

@article{clemons2001estrogen,
  title={Estrogen and the risk of breast cancer},
  author={Clemons, Mark and Goss, Paul},
  journal={New England Journal of Medicine},
  volume={344},
  number={4},
  pages={276--285},
  year={2001},
  publisher={Mass Medical Soc}
}

@article{lupulescu1995clinical,
  title={Clinical science review: estrogen use and cancer incidence: a review},
  author={Lupulescu, Aurel},
  journal={Cancer investigation},
  volume={13},
  number={3},
  pages={287--295},
  year={1995},
  publisher={Taylor \& Francis}
}

@article{anderson2010male,
  title={Male breast cancer: a population-based comparison with female breast cancer},
  author={Anderson, William F and Jatoi, Ismail and Tse, Julia and Rosenberg, Philip S},
  journal={Journal of Clinical Oncology},
  volume={28},
  number={2},
  pages={232--239},
  year={2010},
  publisher={American Society of Clinical Oncology}
}

@article{gleason2012breast,
  title={Breast cancer incidence in black and white women stratified by estrogen and progesterone receptor statuses},
  author={Gleason, Michael X and Mdzinarishvili, Tengiz and Sherman, Simon},
  journal={PLoS One},
  volume={7},
  number={11},
  pages={e49359},
  year={2012},
  publisher={Public Library of Science San Francisco, USA}
}

@article{armitage2004age,
  title={The age distribution of cancer and a multi-stage theory of carcinogenesis},
  author={Armitage, Peter and Doll, Richard},
  journal={British journal of cancer},
  volume={91},
  number={12},
  pages={1983--1989},
  year={2004},
  publisher={Nature Publishing Group}
}

@article{moolgavkar1979two,
  title={Two-event models for carcinogenesis: incidence curves for childhood and adult tumors},
  author={Moolgavkar, Suresh H and Venzon, David J},
  journal={Mathematical biosciences},
  volume={47},
  number={1-2},
  pages={55--77},
  year={1979},
  publisher={Elsevier}
}

@article{moolgavkar1981mutation,
  title={Mutation and cancer: a model for human carcinogenesis},
  author={Moolgavkar, Suresh H and Knudson, Alfred G},
  journal={JNCI: Journal of the National Cancer Institute},
  volume={66},
  number={6},
  pages={1037--1052},
  year={1981},
  publisher={Oxford University Press}
}

@article{brouwer2018case,
  title={Case studies of gastric, lung, and oral cancer connect etiologic agent prevalence to cancer incidence},
  author={Brouwer, Andrew F and Eisenberg, Marisa C and Meza, Rafael},
  journal={Cancer research},
  volume={78},
  number={12},
  pages={3386--3396},
  year={2018},
  publisher={American Association for Cancer Research}
}

@article{mohammad2025modeling,
  title={Modeling Early-Onset Cancer Kinetics Reveals Changes in Underlying Risk and the Impact of Population Screening},
  author={Mohammad Mirzaei, Navid and Hur, Chin and Terry, Mary Beth and Dalerba, Piero and Yang, Wan},
  journal={Cancer Research},
  year={2025},
  publisher={American Association for Cancer Research}
}

@article{meza2008analysis,
  title={Analysis of lung cancer incidence in the nurses’ health and the health professionals’ follow-up studies using a multistage carcinogenesis model},
  author={Meza, Rafael and Hazelton, William D and Colditz, Graham A and Moolgavkar, Suresh H},
  journal={Cancer causes \& control},
  volume={19},
  number={3},
  pages={317--328},
  year={2008},
  publisher={Springer}
}

@misc{Laronningen2025,
  author       = {Larønningen, S. and Arvidsson, G. and Bray, F. and Dahl-Olsen, E. D. and Engholm, G. and Ervik, M. and Friis, S. and Guðmundsdóttir, E. M. and Gulbrandsen, J. and Hansen, H. M. and Johannesen, T. B. and Kristensen, S. and Kønig, S. M. and Lam, F. and Laversanne, M. and Lydersen, L. N. and Malila, N. and Mangrud, O. M. and Miettinen, J. and Pejicic, S. and Persson, Å. and Pettersson, D. and Skog, A. and Steig, B. Á. and Tian, H. and Aagnes, B. and Storm, H. H.},
  title        = {{NORDCAN: Cancer Incidence, Mortality, Prevalence and Survival in the Nordic Countries, Version 9.5 (19.06.2025)}},
  year         = {2025},
  howpublished = {Association of the Nordic Cancer Registries. Cancer Registry of Norway},
  note         = {Available from: \url{https://nordcan.iarc.fr/}, accessed on 09/09/2025},
}

@article{Engholm2010,
  author    = {Engholm, G. and Ferlay, J. and Christensen, N. and Bray, F. and Gjerstorff, M. L. and Klint, A. and Køtlum, J. E. and Olafsdóttir, E. and Pukkala, E. and Storm, H. H.},
  title     = {NORDCAN – a Nordic tool for cancer information, planning, quality control and research},
  journal   = {Acta Oncologica},
  year      = {2010},
  volume    = {49},
  number    = {5},
  pages     = {725--736},
  doi       = {10.3109/02841861003782017},
  pmid      = {20491528},
  url       = {https://doi.org/10.3109/02841861003782017}
}

@misc{NCI,
  title        = {{The Surveillance, Epidemiology, and End Results (SEER) Program}},
  year         = {2025},
  howpublished = {National Cancer Institute},
  note         = {Available from: \url{https://seer.cancer.gov}, accessed on 06/20/2024},
}

@article{tomasetti2015only,
  title={Only three driver gene mutations are required for the development of lung and colorectal cancers},
  author={Tomasetti, Cristian and Marchionni, Luigi and Nowak, Martin A and Parmigiani, Giovanni and Vogelstein, Bert},
  journal={Proceedings of the National Academy of Sciences},
  volume={112},
  number={1},
  pages={118--123},
  year={2015},
  publisher={National Academy of Sciences}
}

@misc{coddington1956theory,
  title={Theory of ordinary differential equations},
  author={Coddington, Earl A and Levinson, Norman and Teichmann, T},
  year={1956},
  publisher={American Institute of Physics}
}

@misc{NCHS2021,
  author       = {{National Center for Health Statistics}},
  title        = {Health, United States, 2020–2021: Table canbrtest},
  howpublished = {\url{https://www.cdc.gov/nchs/hus/data-finder.htm}},
  note         = {Accessed on 11/05/2025}
}

@article{mirzaei2025estimating,
  title={Estimating the carcinogenesis timelines in early-onset versus late-onset cancers and changes across birth cohorts},
  author={Mirzaei, Navid Mohammad and Yang, Wan},
  journal={medRxiv},
  pages={2025--08},
  year={2025},
  publisher={Cold Spring Harbor Laboratory Press}
}

@article{gerstung2020evolutionary,
  title={The evolutionary history of 2,658 cancers},
  author={Gerstung, Moritz and Jolly, Clemency and Leshchiner, Ignaty and Dentro, Stefan C and Gonzalez, Santiago and Rosebrock, Daniel and Mitchell, Thomas J and Rubanova, Yulia and Anur, Pavana and Yu, Kaixian and others},
  journal={Nature},
  volume={578},
  number={7793},
  pages={122--128},
  year={2020},
  publisher={Nature Publishing Group UK London}
}

@article{anderson2014many,
  title={How many etiological subtypes of breast cancer: two, three, four, or more?},
  author={Anderson, William F and Rosenberg, Philip S and Prat, Aleix and Perou, Charles M and Sherman, Mark E},
  journal={Journal of the National Cancer Institute},
  volume={106},
  number={8},
  pages={dju165},
  year={2014},
  publisher={Oxford University Press US}
}

@article{abubakar2023host,
  title={Host, reproductive, and lifestyle factors in relation to quantitative histologic metrics of the normal breast},
  author={Abubakar, Mustapha and Klein, Alyssa and Fan, Shaoqi and Lawrence, Scott and Mutreja, Karun and Henry, Jill E and Pfeiffer, Ruth M and Duggan, Maire A and Gierach, Gretchen L},
  journal={Breast Cancer Research},
  volume={25},
  number={1},
  pages={97},
  year={2023},
  publisher={Springer}
}

@article{zhao2020cancer,
  title={Cancer-associated adipocytes: emerging supporters in breast cancer},
  author={Zhao, Chongru and Wu, Min and Zeng, Ning and Xiong, Mingchen and Hu, Weijie and Lv, Wenchang and Yi, Yi and Zhang, Qi and Wu, Yiping},
  journal={Journal of Experimental \& Clinical Cancer Research},
  volume={39},
  number={1},
  pages={156},
  year={2020},
  publisher={Springer}
}

@article{fu2018foxp1,
  title={Foxp1 is indispensable for ductal morphogenesis and controls the exit of mammary stem cells from quiescence},
  author={Fu, Nai Yang and Pal, Bhupinder and Chen, Yunshun and Jackling, Felicity C and Milevskiy, Michael and Vaillant, Fran{\c{c}}ois and Capaldo, Bianca D and Guo, Fusheng and Liu, Kevin H and Rios, Anne C and others},
  journal={Developmental Cell},
  volume={47},
  number={5},
  pages={629--644},
  year={2018},
  publisher={Elsevier}
}

@article{hecht2016role,
  title={The role of oxidative stress on breast cancer development and therapy},
  author={Hecht, Fabio and Pessoa, Carolina F and Gentile, Luciana B and Rosenthal, Doris and Carvalho, Denise P and Fortunato, Rodrigo S},
  journal={Tumor biology},
  volume={37},
  number={4},
  pages={4281--4291},
  year={2016},
  publisher={Springer}
}

@article{deng2022biological,
  title={Biological role of matrix stiffness in tumor growth and treatment},
  author={Deng, Boer and Zhao, Ziyi and Kong, Weimin and Han, Chao and Shen, Xiaochang and Zhou, Chunxiao},
  journal={Journal of translational medicine},
  volume={20},
  number={1},
  pages={540},
  year={2022},
  publisher={Springer}
}

@article{amens2021immune,
  title={Immune system effects on breast cancer},
  author={Amens, Jensen N and Bah{\c{c}}ecioglu, G{\"o}khan and Zorlutuna, Pinar},
  journal={Cellular and Molecular Bioengineering},
  volume={14},
  number={4},
  pages={279--292},
  year={2021},
  publisher={Springer}
}

@article{li2018mutation,
  title={Mutation mechanisms of human breast cancer},
  author={Li, Lingling and Tian, Tianhai and Zhang, Xinan},
  journal={Journal of Computational Biology},
  volume={25},
  number={4},
  pages={396--404},
  year={2018},
  publisher={Mary Ann Liebert, Inc. 140 Huguenot Street, 3rd Floor New Rochelle, NY 10801 USA}
}

@article{zhang2005estimating,
  title={Estimating the number of rate limiting genomic changes for human breast cancer},
  author={Zhang, Xinan and Simon, Richard},
  journal={Breast cancer research and treatment},
  volume={91},
  number={2},
  pages={121--124},
  year={2005},
  publisher={Springer}
}

\end{document}